\documentclass[aps,pra,twocolumn,groupedaddress]{revtex4-1}
 \pdfoutput=1
\usepackage{amsmath}
\usepackage{graphicx}
\usepackage{amsfonts}
\usepackage{color}
\usepackage{upgreek}
\makeatletter
\def\captionof#1#2{{\def\@captype{#1}#2}}
\makeatother
\usepackage{hyperref}
\hypersetup{
    colorlinks = true,
    linkcolor =blue,
	citecolor=blue, 
	urlcolor=blue 
}

\begin{document}

\title{Classifying transport behavior via current fluctuations in open quantum systems}
\author{Archak Purkayastha}
\affiliation{School of Physics, Trinity College Dublin, College Green, Dublin 2, Ireland}
  
\date{\today}

\begin{abstract}
There are two standard ways of classifying transport behavior of systems. The first is via time scaling of spread of correlations in the isolated system in thermodynamic limit. The second is via system size scaling of conductance in the steady state of the open system. We show here that these correspond to taking the thermodynamic limit and the long time limit of the integrated equilibrium current-current correlations of the open system in different order. In general, the limits may not commute leading to a conflict between the two standard ways of transport classification. Nevertheless, the full information is contained in the equilibrium current-current correlations of the open system. We show this  analytically by rigorously deriving the open-system current fluctuation dissipation relations (OCFDR) starting from an extremely general open quantum set-up and then carefully taking the proper limits. We test our theory numerically on the non-trivial example of the critical Aubry-Andr{\'e}-Harper (AAH) model, where, it has been recently shown that, the two standard classifications indeed give different results. We find that both the total current autocorrelation and the long-range local current correlations of the open system in equilibrium show signatures of diffusive transport up to a time scale. This time scale grows as square of system size. Beyond this time scale a steady state value is reached. The steady state value is conductance, which shows sub-diffusive scaling with system size.
\end{abstract}

\maketitle

\subsection{Introduction}

Fluctuation-dissipation theorem is one of the fundamental concepts of physics, and is of interest across all of physics. In terms of current fluctuations in isolated systems in the thermodynamic limit, it manifests as the standard isolated system Green-Kubo formula  \cite{Kubo1,Kubo2,Green1}. The isolated system Green-Kubo formula describes the linear response of an isolated macroscopic system to some internal gradient assumed to be set-up due to some external temperature or chemical potential bias. Transport coefficients like particle conductivity can be calculated from the isolated system Green-Kubo formula. Under certain standard assumptions, these can also be related to spread of correlations (density correlations in case of particle conductivity) in the isolated system in the thermodynamic limit. The time scaling of the spread of correlations shows whether the corresponding transport coefficient is finite or zero or diverging and how these limits are approached. This makes it possible to classify transport behavior of the system in terms of time scaling of spread of correlations. This is one of the standard ways of classifying transport behavior (into diffusive, sub-diffusive, localized,  super-diffusive or ballistic). We call this the isolated system classification of transport. 

Alternatively, one can connect the system to two baths (leads) at two ends and calculate  (measure) the steady state conductance. The scaling of conductance with the length of the system (in between the two baths) shows whether the corresponding conductivity is finite or zero or diverging in the thermodynamic limit and also how these limits are approached. This gives the second standard way of classifying transport behaviors (again into diffusive, sub-diffusive, localized,  super-diffusive or ballistic). We call this the open system classification of transport.

Usually, the above two standard ways of classifying transport behaviors are consistent, but this may not be always so \cite{oc_mismatch1,oc_mismatch2,oc_mismatch3}. In fact, recent studies in open and isolated quasiperiodic systems have shown that the two standard ways of classification can give drastically different results ~\cite{oc_mismatch3, vkv}. The particular case in point is the Aubry-Andr{\'e}-Harper (AAH) model \cite{aa1,harper}. This is a model of non-interacting particles in a one dimensional lattice in presence of a quasiperiodic potential. Upon increasing the strength of the potential, there occurs a phase transition from all eigenstates being completely delocalized to all eigenstates being completely localized. The phase transition is mediated by a critical point where the eigenstates are neither completely delocalized nor localized but are `critical' \cite{pandit83} and the spectrum has fractal properties \cite{TangKohmoto1986}.  In Ref.~\cite{oc_mismatch3}, it has been shown that transport in the critical Aubry-Andr{\'e}-Harper (AAH) model is `diffusive-like' according to the isolated system classification, while it is sub-diffusive according to the open system classification. (We say `diffusive-like' because, that isolated system transport is not strictly diffusive.  It was shown in Ref.~\cite{oc_mismatch3} that there exists some hints of super-diffusive behavior in the isolated system. For strictly diffusive systems, the diffusion constant calculated from spread of correlations in the isolated system and that from the scaling of conductance of the open system are expected to match \cite{znidaric_Kubo_2018}.)

The above results suggest that we need to revisit our understanding of the classification of transport behavior. Particularly, we need to find the connection between the two standard ways of transport classification. In this paper, we show that both the isolated system classification and the open system classification actually probe the equilibrium current-current correlations of the open system, but under different time and length scales. In other words, in cases where they give different results, (for example, that described above) both behaviors will be seen in the equilibrium current fluctuations of the open system.

\begin{figure}
\includegraphics[width=\columnwidth]{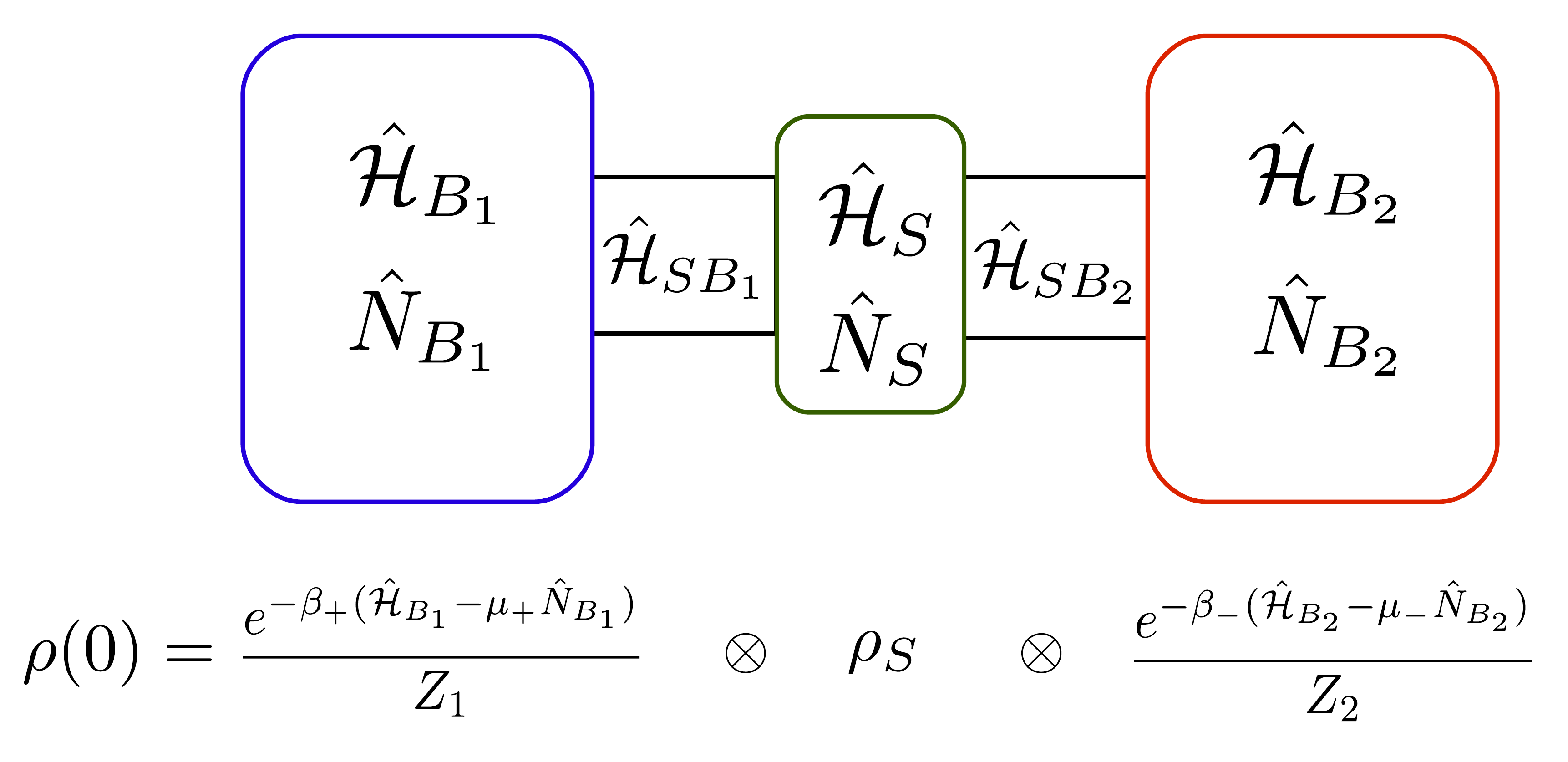}
\caption{(color online)  Our general set-up. $H_S$ is the system Hamiltonian, $H_{B_1}$ ($H_{B_2}$) is the left (right) bath Hamiltonian, $H_{SB_1}$ ($H_{SB_2}$) is the Hamiltonian coupling system to left (right) bath. $N_S, N_{B_1}, N_{B_2}$ are operators corresponding to total number of particles in system, left bath and right bath. The baths have infinite degrees of freedom. The initial state of the set-up is $\rho(0)$. Initially, the baths are at their respective thermal states with slightly different (inverse) temperatures and chemical potentials $\beta_{\pm}=\beta \pm \frac{\Delta \beta}{2}$, $\mu_{\pm}=\mu \pm \frac{\Delta \mu}{2}$. The system is initially at some arbitrary state $\rho_S$. At $t\rightarrow\infty$, the system reaches a NESS showing transport (i.e., having non-zero currents). We are interested in linear response  particle and energy  transport coefficients when $\Delta \beta,\Delta \mu \rightarrow 0$.}
\label{fig:set_up}
\end{figure}

The standard isolated system Green-Kubo formula gives the current fluctuation-dissipation relation for the isolated system. So, to bring the open quantum system problem to the same footing,  we need to find the open-system current fluctuation-dissipation relations (OCFDR). While there are expected answers to this based on experiments \cite{Johnson,Nyquist,Kubo_expt} and previous investigations in non-interacting quantum systems \cite{FisherLee, shot_noise, Abhishek2}, we would like a rigorous and more general derivation. In the first part of the paper, we give our derivation of the OCFDR under very general conditions starting from the set-up of an arbitrary system connected to two arbitrary baths with slightly different temperatures and chemical potentials (see Fig.~\ref{fig:set_up}). The only assumptions we make are time-translational and time-reversal invariance of the full system+bath Hamiltonian, open system thermalization and the so-called mixing assumption (to be explained below). There has been several attempts to generalize linear response theory to open quantum systems \cite{Ban2017,Shen2017,Zanardi2016,Shi2016,Ban2015_1,Ban2015_2,Avron2012,Chetrite2012,
 Avron2011, Uchiyama2010, Uchiyama2009, FisherLee, DaviesSpohn, Kamiya2013, Wu2010, Jaksic2007, Jaksic2006_1, Jaksic2006_2,Jaksic2006_3}. But, even after extensive literature survey, no reference could be found where exactly this derivation for the open quantum system has been given in as much generality. Further, we obtain a number of new fundamental results. For the corresponding classical problem, however, the current fluctuation-dissipation relations  were obtained in full generality in Refs.~\cite{Onuttam2011,Anupam2011,Anupam2009}.  
 
In deriving the OCFDR, we first find a linear response expression for the non-equilibrium steady state (NESS) density matrix without any further approximations (arbitrary system, bath and system-bath coupling Hamiltonians, arbitrary system-size, no weak system-bath coupling, no Markovian assumption). This very non-trivial result shows that set-ups which show thermalization will always relax to a unique NESS in the linear response regime, irrespective of the initial state of the system. Then we derive the OCFDR as expressions for the elements of the Onsager matrix \cite{Onsager1,Onsager2} for thermoelectric transport coefficients. These expressions show, in general, the Onsager relation, written in terms of system currents, can be violated. The Onsager relation is recovered under the assumption of short-ranged system Hamiltonian. Thus, our result also gives the form of the Onsager matrix for system Hamiltonians having long range terms, where the Onsager relation may not hold. Further, for short-ranged systems, we find the rather surprising result that the time integrated current-current correlation between \emph{any two local currents} of the open system in equilibrium is the \emph{same} and is proportional to the corresponding transport coefficient. This is in stark contrast with the isolated system Green-Kubo formula which involves only total currents of the system.  Our results for OCFDR for short-ranged system generalizes previously known results for non-interacting systems \cite{FisherLee} to interacting systems under much more realistic assumptions.

Next we show that, for short-ranged systems, the integrated total current-current correlations of the open system shows a crossover between open-system behavior and isolated thermodynamic limit behavior as a function of system-size and observation time, even with strong system-bath coupling. As a consequence, the isolated system and the open system classifications of transport correspond to taking the thermodynamic limit and the long time limit of the integrated total current-current correlations of the open system in different order. In general the limits may not commute leading to conflicting isolated system and open system classifications. Finally, we work out the non-trivial example of the open critical AAH model where actually the limits to not commute. We show that the diffusive-like behavior persists in current fluctuations up to some time scale. This scale grows as square of system size. Beyond this time scale a steady state is reached, which is sub-diffusive. We also find that the long range correlations between far-off local currents of the open system in equilibrium also shows signatures of both the diffusive-like behavior of the isolated system and the sub-diffusive behavior of the open system NESS. To our knowledge, this is the first work explicitly showing that time and system-size dependence of current fluctuations of an open system in equilibrium can be used to classify both isolated system and open system transport behaviors of a model.

The paper is organized as follows. In Section~\ref{set-up}, we give the details of the set-up and the basic assumptions required. In Section~\ref{NESS}, we obtain the general expression for NESS density matrix in linear response regime. In Section~\ref{OCFDR}, we give the OCFDR, which, in general, violate the Onsager relation. In Section~\ref{Onsager}, under the assumption of short-ranged system Hamiltonian, we give various other equivalent forms of the OCFDR where Onsager relation is recovered.  In Section~\ref{open-closed}, the crossover of integrated equilibrium current-current correlations between open system and isolated system behaviors is derived. In Section~\ref{consequences}, we discuss the consequences of this crossover for the standard ways of classification of transport. In Section~\ref{AAH}, we work out the non-trivial example of the critical AAH model. In Section~\ref{conclusions}, we summarize our results point by point and give the future research directions. The reader may choose to read Section~\ref{conclusions} first to get to know the gist of the main results, without going through the details.  Throughout the manuscript, $\hbar$ has been set to $1$.

\subsection{The set-up, definitions and assumptions}\label{set-up}

We take a system that is connected to two baths at its two ends (Fig.~\ref{fig:set_up}). The full system+baths Hamiltonian is given by
\begin{align}
\hat{\mathcal{H}} = \hat{\mathcal{H}}_S + \hat{\mathcal{H}}_{SB_1} + \hat{\mathcal{H}}_{B_1} + \hat{\mathcal{H}}_{SB_2} + \hat{\mathcal{H}}_{B_2},
\end{align}
where $\hat{\mathcal{H}}_S$ is system Hamiltonian, $\hat{\mathcal{H}}_{B_1}$ ($\hat{\mathcal{H}}_{B_2}$) is the left (right) bath Hamiltonian,  $\hat{\mathcal{H}}_{SB_1}$ ($\hat{\mathcal{H}}_{SB_2}$) is system-bath coupling Hamiltonian for left (right) bath. We assume system and bath Hamiltonians to be number conserving. So $[\hat{N}_S,\hat{\mathcal{H}}_S]=[\hat{N}_{B_1}, \hat{\mathcal{H}}_{B_1}]=[\hat{N}_{B_2}, \hat{\mathcal{H}}_{B_2}]=0$, where $\hat{N}_S, \hat{N}_{B_1}, \hat{N}_{B_2}$ are the total number operators of system, left bath and right bath respectively. We will also assume $ [\hat{N}_S + \hat{N}_{B_p},\hat{\mathcal{H}}_{SB_p}] = 0$,  $p=1,2$. This physically means that the system-bath coupling Hamiltonians do not act as `sources' of particle. In this set-up, we define the following current operators:
\begin{align}
\label{curr_def}
\hat{I}_{B_p\rightarrow S} &= -\frac{d\hat{N}_{B_p}}{dt}= i[\hat{N}_{B_p},\hat{\mathcal{H}}_{SB_p}] \nonumber \\
&=-\hat{I}_{S \rightarrow B_p},  \nonumber\\
\hat{J}_{B_p\rightarrow S} &=-\frac{d\hat{\mathcal{H}}_{B_p}}{dt}= i[\hat{\mathcal{H}}_{B_p},\hat{\mathcal{H}}_{SB_p}] \nonumber \\
&=-\hat{J}_{S \rightarrow B_p},  
\end{align}
$p=1,2$. The first line gives particle currents between the baths and the system. The second line gives energy currents between the baths and the system. We also define the operators $\hat{I}_S$ and $\hat{J}_S$ as the total particle and energy current operators of the system. We will look at the OCFDR for $\hat{I}_S$, $\hat{J}_S$. This corresponds to the transport coefficients. We assume that each of the system, bath and system-bath coupling Hamiltonians has \emph{time reversal} and \emph{time translation} symmetries.

Let us also define the following notations:
\begin{align}
\label{rho_notations}
&\rho_{EIS}^{\hat{\mathcal{H}}} \equiv \frac{e^{-\beta(\hat{\mathcal{H}}_{B_1}-\mu \hat{N}_{B_1})}}{Z_1}\otimes \rho_S \otimes \frac{e^{-\beta(\hat{\mathcal{H}}_{B_2}-\mu \hat{N}_{B_2})}}{Z_2} \nonumber \\
&\rho_{EIS}^{\hat{\mathcal{H}}}(\infty) \equiv \lim_{t\rightarrow \infty} e^{-i{\hat{\mathcal{H}}}t}\rho_{EIS}^{\hat{\mathcal{H}}} e^{i{\hat{\mathcal{H}}}t} \nonumber \\
& \rho_{eq}^{\mathcal{\hat{H}}} \equiv \frac{e^{-\beta({\hat{\mathcal{H}}}-\mu \hat{N})}}{Z}, \nonumber \\
& \langle ... \rangle = Tr(... \rho_{eq}^{\mathcal{\hat{H}}}) 
\end{align}
where $\hat{N}=\hat{N}_S+\hat{N}_{B_1}+\hat{N}_{B_2}$, $\rho_S$ is some arbitrary initial state of the system and $Z_1$, $Z_2$, $Z$ are corresponding normalization constants (partition functions). The superscripts denote that time evolution is with the Hamiltonian $\hat{\mathcal{H}}$. 
Armed with the above definitions, now, we make the most crucial assumption of open system thermalization.  We say that the set-up shows \emph{open system thermalization} if the following holds
\begin{align}
\label{thermalization}
& Tr(\hat{P}\rho_{EIS}^{\hat{\mathcal{H}}}(\infty))=Tr(\hat{P}\rho_{eq}^{\mathcal{\hat{H}}})= \langle \hat{P} \rangle, \nonumber \\
& Tr(e^{i\mathcal{\hat{H}}t}\hat{P}e^{-i\mathcal{\hat{H}}t}\hat{Q}\rho_{EIS}^{\hat{\mathcal{H}}}(\infty))=Tr(e^{i\mathcal{\hat{H}}t}\hat{P}e^{-i\mathcal{\hat{H}}t}\hat{Q}\rho_{eq}^{\mathcal{\hat{H}}}) \nonumber \\
&= \langle \hat{P}(t)\hat{Q}(0) \rangle,
\end{align}
where $\hat{P}$ and $\hat{Q}$ are either any two system operators or the current opetarors from the baths defined in Eq.~\ref{curr_def}. In words, the above equations mean that when the system is connected to two baths at same temperature and chemical potential, the system observables in long time limit behave as if the full system+bath set-up has reached the thermal state with the same temperature and chemical potential, irrespective of the initial state of the system. While this is physically expected to be generically true, the conditions for validity of the above assumption are not known in general. But, for non-interacting systems (i.e, where the full system+bath Hamiltonian $\mathcal{\hat{H}}$ is quadratic), following Refs.~\cite{dharroy2006,dharsen2006}, it can be shown to hold if: a) the spectral functions of the baths are well approximated by continuous functions, which in turn requires that the baths have infinite of degrees of freedom, b) the bandwidths of the baths are larger than that of the system. On physical grounds, without proof, we will assume that our  set-up is such that Eq.~\ref{thermalization} also holds true for interacting systems. Note that, it is the choice of the initial state $\rho_{EIS}^{\hat{\mathcal{H}}}$ that clearly demarcates the system, the baths and the system bath coupling Hamiltonians.  (If the reader is familiar with the isolated system eigenstate thermalization hypothesis (ETH) statement \cite{ETH_review}, we point to Appendix~\ref{ETH} for a discussion.)   

Finally, we define the following notation
\begin{align}
\label{notation}
& M(\hat{Q},\hat{P}) \equiv \frac{1}{\beta}\int_0^\infty dt \int_0^\beta d\lambda \langle \hat{Q}(- i\lambda) \hat{P}(t) \rangle \nonumber\\
& = \frac{1}{\beta}\lim_{\tau\rightarrow \infty}\int_0^\tau dt \int_0^\beta d\lambda \langle \hat{Q}(- i\lambda) \hat{P}(t) \rangle \nonumber \\
\end{align}
Using time translation and time-reversal symmetries, one can also show $M(\hat{Q},\hat{P})=M(\hat{P},\hat{Q})$ (see Appendix~\ref{MqpMpq}). If $[\hat{Q},\hat{N}]=0$, assuming the limit exists, there are no singularities of $\langle\hat{Q}(0) \hat{P}(z)\rangle$  in the upper complex plane and time-reversal and time-translation invariances, $M(\hat{Q},\hat{P})$ can be reduced to (see Appendix~\ref{simpf_Mqp})
\begin{align}
\label{notation2}
& M(\hat{Q},\hat{P}) =  \frac{1}{2}\int_{-\infty}^\infty dt \langle \hat{Q}(t) \hat{P}(0)\rangle. 
\end{align}

\subsection{The linear response NESS}\label{NESS}
We are interested in linear response, so let  $\beta_{\pm} = \beta\pm\epsilon\Delta\beta/2$, $\mu_{\pm} = \mu\pm\epsilon\Delta\mu/2$, $\epsilon \rightarrow 0$. We start the set-up in the following non-equilibrium initial state $\rho_{NIS}^{\hat{\mathcal{H}}}$ (see Fig.~\ref{fig:set_up}),
\begin{align}
\label{non-eq_intitial}
&\rho(0)=\rho_{NIS}^{\hat{\mathcal{H}}}, \nonumber \\ &\rho_{NIS}^{\hat{\mathcal{H}}}\equiv\frac{e^{-\beta_+({\hat{\mathcal{H}}}_{B_1}-\mu_+ \hat{N}_{B_1})}}{Z_1}\otimes \rho_S \otimes \frac{e^{-\beta_-({\hat{\mathcal{H}}}_{B_2}-\mu_- \hat{N}_{B_2})}}{Z_2} \nonumber\\
&= \frac{e^{-\beta({\hat{\mathcal{H}}}_{B_1}^{\prime}-\mu \hat{N}_{B_1})}}{Z_1}\otimes \rho_S \otimes \frac{e^{-\beta({\hat{\mathcal{H}}}_{B_2}^{\prime}-\mu \hat{N}_{B_2})}}{Z_2},
\end{align}
with  
\begin{align}
&{\hat{\mathcal{H}}}_{B_1}^{\prime} ={\hat{\mathcal{H}}}_{B_1}+\frac{\epsilon}{2\beta}( {\hat{\mathcal{H}}}_{B_1}\Delta \beta -  \hat{N}_{B_1}\Delta (\mu \beta)), \nonumber \\
& {\hat{\mathcal{H}}}_{B_2}^{\prime} ={\hat{\mathcal{H}}}_{B_2}-\frac{\epsilon}{2\beta}( {\hat{\mathcal{H}}}_{B_2}\Delta \beta -  \hat{N}_{B_2}\Delta (\mu \beta))
\end{align}
and $\Delta(\mu \beta)=\beta \Delta \mu+\mu \Delta \beta$.  $\rho_{NIS}^{\hat{\mathcal{H}}}$ is the standard initial condition used to obtain NESS results in open quantum systems. In this case also, the choice of the initial state $\rho_{NIS}^{\hat{\mathcal{H}}}$ clearly demarcates the system, the baths and the system bath coupling Hamiltonians.  In obtaining the second line of Eq.~\ref{non-eq_intitial}, we have only regrouped the terms in the exponentials and neglected the $\epsilon^2$ term. We define 
\begin{align}
{\hat{\mathcal{H}}}^{\prime} \equiv {\hat{\mathcal{H}}}_S + {\hat{\mathcal{H}}}_{SB_1}+{\hat{\mathcal{H}}}_{B_1}^{\prime}+{\hat{\mathcal{H}}}_{SB_2}
+{\hat{\mathcal{H}}}_{B_2}^{\prime}={\hat{\mathcal{H}}}+\epsilon {\hat{\mathcal{H}}}_{pert},
\end{align}
where 
\begin{align}
\label{H_perturb}
{\hat{\mathcal{H}}}_{pert} = \frac{1}{\beta}[\Delta \beta(\frac{{\hat{\mathcal{H}}}_{B_1} - {\hat{\mathcal{H}}}_{B_2}}{2}) + \Delta(-\beta \mu)(\frac{\hat{N}_{B_1} - \hat{N}_{B_2}}{2})],
\end{align}
Comparing the second line of Eq.~\ref{non-eq_intitial} with Eq.~\ref{thermalization}, we see that $\rho_{NIS}^{\hat{\mathcal{H}}}=\rho_{EIS}^{{\hat{\mathcal{H}}}^\prime}$.  Thus we make the crucial observation that $\rho_{NIS}^{\hat{\mathcal{H}}}$ is the non-equilibrium initial state when evolved with the Hamiltonian ${\hat{\mathcal{H}}}$, but, when evolved with ${\hat{\mathcal{H}}}^\prime$, it is an equilibrium initial state and reaches $\rho_{EIS}^{\hat{\mathcal{H}}^\prime}(\infty)$ in the long time limit,
\begin{align}
\label{H_prime}
&\lim_{t\rightarrow \infty} e^{-i{\hat{\mathcal{H}}}^\prime t}\rho_{NIS}^{\hat{\mathcal{H}}} e^{i{\hat{\mathcal{H}}}^\prime t} = 
\lim_{t\rightarrow \infty} e^{-i{\hat{\mathcal{H}}}^\prime t}\rho_{EIS}^{{\hat{\mathcal{H}}}^\prime} e^{i{\hat{\mathcal{H}}}^\prime t}=\rho_{EIS}^{\hat{\mathcal{H}}^\prime}(\infty) 
\end{align}
We are interested in time evolution with ${\hat{\mathcal{H}}}$. This is given by, $\frac{\partial\rho}{\partial t}= i [\rho,{\hat{\mathcal{H}}}] = i [\rho,{\hat{\mathcal{H}}}^{\prime}] - i\epsilon [\rho,{\hat{\mathcal{H}}}_{pert}]$. Assuming $\hat{\mathcal{H}}^{\prime}$ as the unperturbed Hamiltonian, solving for $\rho(t)$ upto linear order in $\epsilon$ (linear response) using Dyson series and taking $t\rightarrow \infty$ and using Eq.~\ref{H_prime} (see Appendix~\ref{Dyson}), we have
\begin{align}
\label{rho_NESS}
&\rho_{NESS}^{\hat{\mathcal{H}}} =\lim_{t\rightarrow \infty} \rho(t)\nonumber \\
& = \rho_{EIS}^{\hat{\mathcal{H}}^\prime}(\infty)-i\epsilon \int_0^\infty dt^\prime [\rho_{EIS}^{\hat{\mathcal{H}}}(\infty),  e^{-i{\hat{\mathcal{H}}} t^\prime}{\hat{\mathcal{H}}}_{pert}e^{i{\hat{\mathcal{H}}} t^\prime}] 
\end{align}
In the second term, we have used ${\hat{\mathcal{H}}}^{\prime}\rightarrow {\hat{\mathcal{H}}}$, because corrections above this will be of order $\epsilon^2$. Taking expectation value of any system operator $\hat{O}$, and using time-translation invariance and open system thermalization (Eq.~\ref{thermalization}), we have,
\begin{align}
\label{O_NESS}
&\langle \hat{O} \rangle_{NESS}= \langle \hat{O} \rangle_{\mathcal{H}^\prime}+i\epsilon \int_{0}^{\infty}dt\langle [\hat{O}(t), {\hat{\mathcal{H}}}_{pert}]\rangle \nonumber \\
& = \langle \hat{O} \rangle_{\mathcal{H}^\prime}-\epsilon \big[\Delta \beta~M(\hat{J}_B,\hat{O})+\Delta(-\beta \mu)~M(\hat{I}_B,\hat{O})\big],
\end{align}
where $\langle ... \rangle_{NESS} = Tr(... \rho_{NESS}^{\hat{\mathcal{H}}})$, $\langle ... \rangle_{\mathcal{H}^\prime}=Tr(...\rho_{eq}^{\mathcal{\hat{H}}^\prime})$, $\hat{I}_B(t) = [\hat{I}_{B_1\rightarrow S}(t)+\hat{I}_{S\rightarrow B_2}(t)]/2$, $\hat{J}_B(t) = [\hat{J}_{B_1\rightarrow S}(t)+\hat{J}_{S\rightarrow B_2}(t)]/2$, and $M(\hat{Q},\hat{P})$ is as defined in Eq.~\ref{notation}. Obtaining the second line from the first line requires some algebra, given in Appendix~\ref{Kubo_trick}.    
  \textit{Eq.~\ref{O_NESS} shows that, in general set-ups showing open system thermalization (Eq.~\ref{thermalization}), a unique NESS is reached by system observables in linear response regime, irrespective of the initial state of the system.} 
This is a very non-trivial and fundamentally important result regarding NESS of general open quantum systems. To our knowledge, this has not been shown before. 

Note that $\hat{I}_B(t),\hat{J}_B(t)$ are a symmetric combinations of currents from the left bath and currents into the right bath. This is an artefact of choosing the initial inverse temperatures (chemical potentials) of the baths as $\beta\pm\epsilon\Delta\beta/2$ ($\mu\pm\epsilon\Delta\mu/2$). Since $\beta$ ($\mu$) is completely arbitrary and only the difference in the temperatures and chemical potentials between the baths matter, we could have completely equivalently chosen the initial inverse temperatures of the baths as $\beta+\epsilon\Delta \beta$ and $\beta$ ($\mu+\epsilon\Delta \mu$, and $\mu$). In that case, we would have found  $\hat{J}_B(t) = \hat{J}_{B_1\rightarrow S}(t)$ ($\hat{I}_B(t) = \hat{I}_{B_1\rightarrow S}(t)$).  Similarly, by choosing the initial inverse temperatures of the baths as $\beta$ and $\beta-\epsilon\Delta \beta$ ($\mu$ and $\mu-\epsilon\Delta \mu$), we would have found $\hat{J}_B(t) = \hat{J}_{S\rightarrow B_2}(t)$ ($\hat{I}_B(t) = \hat{I}_{S\rightarrow B_2}(t)$). These three cases are physically identical.

Another important point to note is that in calculating $\langle \hat{O} \rangle_{NESS}$ by Eq.~\ref{O_NESS}, if  $\langle \hat{O} \rangle_{\mathcal{H}^\prime}\neq 0$, all orders of $\varepsilon$ are generated. But only upto $O(\varepsilon)$ can be trusted. In the following, we will have $\hat{O}$ as a current operator for which $\langle \hat{O} \rangle_{\mathcal{H}^\prime}=0$. So we will not encounter this issue.

\vspace{15pt}

\subsection{The OCFDR}\label{OCFDR}
 If $\hat{O}$ is an energy or particle current operator, we have $\langle \hat{O} \rangle_{\mathcal{H}^\prime}=0$, because energy and particle current operators are odd under time-reversal while ${\hat{\mathcal{H}}}^\prime$ is even under time reversal.Writing Eq.~\ref{O_NESS} explicitly for $\hat{I}_S$ and $\hat{J}_S$, and omitting $\epsilon$ for notational convenience,  we obtain the transport coefficients
\begin{align}
&\left(
\begin{array}{c}
\langle \hat{J}_S \rangle_{NESS}\\
\langle \hat{I}_S \rangle_{NESS}\\
\end{array}
\right)\equiv 
\left(
\begin{array}{cc}
L_{11} & L_{12}\\
L_{21} & L_{22}\\
\end{array}
\right)
\left(
\begin{array}{c}
\Delta \beta\\
\Delta (-\mu \beta)\\
\end{array}
\right) \nonumber\\
& =
-\left(
\begin{array}{cc}
M(\hat{J}_B,\hat{J}_S) & M(\hat{I}_B,\hat{J}_S) \\
M(\hat{J}_B,\hat{I}_S) & M(\hat{I}_B,\hat{I}_S)\\
\end{array}
\right)
\left(
\begin{array}{c}
\Delta \beta\\
\Delta (-\mu \beta)\\
\end{array}
\right). \label{trasnp_coeff}
\end{align}
where $\hat{I}_B(t) = [\hat{I}_{B_1\rightarrow S}(t)+\hat{I}_{S\rightarrow B_2}(t)]/2$, $\hat{J}_B(t) = [\hat{J}_{B_1\rightarrow S}(t)+\hat{J}_{S\rightarrow B_2}(t)]/2$.  The LHS of above equation involves expectation value of total system currents in NESS under infinitesimal bias, while, the RHS involves expectation value of current fluctuations in the thermal state of the whole system+bath set-up. Thus we have the OCFDR. High temperature limit of Eq.~\ref{trasnp_coeff} reproduces the results for classical Hamiltonian systems connected to Langevin baths \cite{Anupam2009, Onuttam2011}. The result can be straightforwardly generalized to multiple (more than two) baths.

An important point to appreciate regarding open systems is that, transport coefficients of \emph{finite length} systems obtained from a set-up of the type we are considering (Fig.~\ref{fig:set_up}) will always be finite. As a result, the infinite time limits involved in the calculation of RHS of Eq.~\ref{trasnp_coeff} (see Eq.~\ref{notation}) will always exist for \emph{finite length} open systems. This is in unlike similar infinite time limits that occur in calculation of the transport coefficients by Green-Kubo formula for an isolated system in thermodynamic limit,  which may diverge (for example, a ballistic system).

Eq.~\ref{trasnp_coeff} has a form similar to definition of Onsager transport coefficients, but the Onsager relation $L_{12}=L_{21}$ clearly does not hold in general ($M(\hat{I}_B,\hat{J}_S) \neq  M(\hat{J}_B,\hat{I}_S)$). Note that, since $M(\hat{Q},\hat{P})=M(\hat{P},\hat{Q})$, this would not be the case if $\langle \hat{J}_B \rangle_{NESS}$ , $\langle \hat{I}_B \rangle_{NESS}$ were calculated instead. Using this fact, as shown in the following, the Onsager relation can be recovered under the assumption of a \emph{short-ranged system}.

\subsection{OCFDR for short-ranged systems}\label{Onsager}
A \emph{short-ranged system} is one described by a Hamiltonian that can be \emph{broken up into $L$ surfaces} transverse to direction of current flow such that  
\begin{align}
\label{continuity}
&{\hat{\mathcal{H}}}_S = \sum_{\ell=1}^{L} {\hat{\mathfrak{H}}}_\ell,~\hat{N}_S = \sum_{\ell=1}^{L} \hat{n}_\ell,\nonumber\\
& \hat{I}_S = \sum_{\ell=1}^{L-1} \hat{I}_{\ell},~ \hat{J}_S = \sum_{\ell=1}^{L-1} \hat{J}_{\ell}  \\
&\frac{d\hat{n}_\ell}{dt} = \hat{I}_{\ell-1} - \hat{I}_{\ell}, \hspace{15pt} \frac{d\hat{\mathfrak{H}}_\ell}{dt} = \hat{J}_{\ell-1} - \hat{J}_{\ell} \nonumber\\
&\frac{d\hat{n}_1}{dt} = \hat{I}_{B_1 \rightarrow S} - \hat{I}_{1}, \hspace{15pt}  \frac{d\hat{n}_L}{dt} = \hat{I}_{L-1} - \hat{I}_{S \rightarrow B_2} \nonumber\\
&\frac{d\hat{\mathfrak{H}}_1}{dt} = \hat{J}_{B_1 \rightarrow S} - \hat{J}_{1}, \hspace{15pt}  \frac{d\hat{\mathfrak{H}}_L}{dt} = \hat{J}_{L-1} - \hat{J}_{S \rightarrow B_2}  \nonumber 
\end{align}
Here $\hat{\mathfrak{H}}_\ell$  ($\hat{n}_\ell$) is the local energy  (particle number) operator of $\ell$th surface, and $\hat{J}_{\ell}$ ($\hat{I}_{\ell}$) is the local current operator giving energy (particle) flow between $\ell$th and $\ell+1$th surfaces. An example of a short-ranged system is a system with nearest-neighbour interactions and hopping. On the other hand, \emph{long-ranged systems} are ones it is not possible to write Eq.~\ref{continuity}, for example, a system with power-law interaction or hopping. Eq.~\ref{trasnp_coeff} holds for both \emph{long-ranged} and \emph{short-ranged} systems. To our knowledge, this is a completely new result for open quantum systems. In the following, we will simplify Eq.~\ref{trasnp_coeff} assuming \emph{short-ranged systems} to obtain some known or expected forms of the OCFDR. While the following forms of the OCFDR may be expected or known, our derivations starting from Eq.~\ref{trasnp_coeff} will provide a more general and rigorous, and less ad-hoc understanding of them. This will also provide important consistency checks for Eq.~\ref{trasnp_coeff}.

By definition, in the NESS, the LHS of the continuity equaions in Eq.~\ref{continuity}  will be zero on average. This leads us to 
\begin{align}
\label{currents_NESS}
&\langle \hat{I}_B \rangle_{NESS} = \langle \frac{\hat{I}_{B_1 \rightarrow S} + \hat{I}_{S \rightarrow B_2}}{2}\rangle_{NESS} \nonumber \\
&= \langle \hat{I}_{B_1 \rightarrow S} \rangle_{NESS}  = \langle \hat{I}_{\ell} \rangle_{NESS}=\frac{\langle \hat{I}_{S} \rangle_{NESS}}{(L-1)},
\end{align}
 and similarly for energy currents.  Using Eq.~\ref{O_NESS} for $\langle \hat{I}_B \rangle_{NESS}$, $\langle \hat{J}_B \rangle_{NESS}$, we have,
\begin{align}
&\left(
\begin{array}{c}
\langle \hat{J}_S \rangle_{NESS}\\
\langle \hat{I}_S \rangle_{NESS}\\
\end{array}
\right)=
(L-1)
\left(
\begin{array}{c}
\langle \hat{J}_B \rangle_{NESS}\\
\langle \hat{I}_B \rangle_{NESS}\\
\end{array}
\right) \nonumber\\
& = 
-(L-1)\left(
\begin{array}{cc}
M(\hat{J}_B,\hat{J}_B) & M(\hat{I}_B,\hat{J}_B) \\
M(\hat{J}_B,\hat{I}_B) & M(\hat{I}_B,\hat{I}_B)\\
\end{array}
\right)
\left(
\begin{array}{c}
\Delta \beta\\
\Delta (-\mu \beta)\\
\end{array}
\right).\label{simpf_transp_coeff1}
\end{align} 
This is the OCFDR in terms of fluctuations of currents from the baths. This form of OCFDR is expected based on experiments \cite{Johnson,Nyquist,Kubo_expt} and previous investigations in non-interacting quantum systems \cite{FisherLee, shot_noise, Abhishek2}. However, our derivation is a rigorous microscopic derivation of them for a very general case including interacting quantum systems. Note that since $M(\hat{Q},\hat{P})=M(\hat{P},\hat{Q})$, now, the Onsager relation is recovered. 

Using Eq.~\ref{O_NESS} and Eq.~\ref{continuity}, we can also write the OCFDR in terms of local system currents :
\begin{align}
&\left(
\begin{array}{c}
\langle \hat{J}_S \rangle_{NESS}\\
\langle \hat{I}_S \rangle_{NESS}\\
\end{array}
\right)=
(L-1)
\left(
\begin{array}{c}
\langle \hat{J}_\ell \rangle_{NESS}\\
\langle \hat{I}_\ell \rangle_{NESS}\\
\end{array}
\right) \nonumber\\
& = 
-(L-1)\left(
\begin{array}{cc}
M(\hat{J}_B,\hat{J}_\ell) & M(\hat{I}_B,\hat{J}_\ell) \\
M(\hat{J}_B,\hat{I}_\ell) & M(\hat{I}_B,\hat{I}_\ell)\\
\end{array}
\right)
\left(
\begin{array}{c}
\Delta \beta\\
\Delta (-\mu \beta)\\
\end{array}
\right).\label{simpf_transp_coeff_local}
\end{align} 
Till now, in all the forms of the OCDFR (Eqs.~\ref{trasnp_coeff},~\ref{simpf_transp_coeff1},~\ref{simpf_transp_coeff_local}), the expressions for the Onsager coefficients involve currents from the baths. If we want to obtain expressions for the Onsager coefficients in terms of equilibrium current fluctuations of the system, without involving currents from the baths, then we need to make a further assumption. 
We call this the \emph{mixing assumption for local currents and densities}, which states the following,
\begin{align}
\label{mixing}
&\lim_{\tau\rightarrow \infty}\langle \hat{n}_m(\pm \tau)\hat{I}_{\ell}(0) \rangle = \lim_{\tau\rightarrow\infty}\langle \hat{n}_m(\pm \tau)\rangle\langle\hat{I}_{\ell}(0) \rangle= 0, \nonumber \\
&\lim_{\tau\rightarrow \infty}\langle \hat{n}_m(\pm \tau)\hat{J}_{\ell}(0) \rangle = \lim_{\tau\rightarrow\infty}\langle \hat{n}_m(\pm \tau)\rangle\langle\hat{J}_{\ell}(0) \rangle= 0 \nonumber \\
&\lim_{\tau\rightarrow \infty}\langle \hat{\mathfrak{H}}_m(\pm \tau)\hat{I}_{\ell}(0) \rangle = \lim_{\tau\rightarrow\infty}\langle \hat{\mathfrak{H}}_m(\pm \tau)\rangle\langle\hat{I}_{\ell}(0) \rangle= 0 \nonumber \\
&\lim_{\tau\rightarrow \infty}\langle \hat{\mathfrak{H}}_m(\pm \tau)\hat{J}_{\ell}(0) \rangle = \lim_{\tau\rightarrow\infty}\langle \hat{\mathfrak{H}}_m(\pm \tau)\rangle\langle\hat{J}_{\ell}(0) \rangle= 0 \nonumber \\
&~~~~\forall~~1\leq m,\ell \leq L.
\end{align}
Barring some pathological cases (such as where, somehow, $\hat{n}_m$, $\hat{I}_\ell$, $\hat{\mathfrak{H}}_m$ or $\hat{J}_\ell$ is a conserved quantity of the whole system+bath set-up), this is generically expected. This is because dissipation due to the infinitely large baths will destroy long-time correlations between system operators. This is consistent with the fact that the set-up shows open system thermalization (Eq.~\ref{thermalization}). To show open system thermalization, dissipation due to the baths must cause long-time correlations between system operators to decay so that the information about the initial state of the system is lost. Note that, the decay need not be exponential (which would be required for a Markovian assumption), but can be a power-law also (which is the typical non-Markovian behavior).

Now,  we can use the same trick as used in Refs.~\cite{Anupam2009, Onuttam2011} for classical systems. In the following, we only consider particle currents. Exactly similar analysis is possible for energy currents.  We define the quantity, $\hat{\mathcal{D}}_m^{n} \equiv \sum_{\ell=1}^m \hat{n}_{\ell} - \sum_{\ell=m}^L \hat{n}_{\ell}$. Taking time derivative using Eq.~\ref{continuity}, we have
\begin{align}
\label{D1}
&\frac{d \hat{\mathcal{D}}_m^{n}}{dt} = 2(\hat{I}_B(t) - \hat{I}_{m}(t))\nonumber \\
& \Rightarrow \hat{\mathcal{D}}_m^{n}(\tau) - \hat{\mathcal{D}}_m^{n}(-\tau)
= 2\int_{-\tau}^{\tau} dt \Big( \hat{I}_B(t) - \hat{I}_{m}(t) \Big)
\end{align}
Multiplying on the right by $\hat{I}_{\ell}(0)$ and taking expectation value, we have,
\begin{align}
\label{Dmt}
&\langle \hat{\mathcal{D}}_m^{n}(\tau)\hat{I}_{\ell}(0) \rangle - \langle \hat{\mathcal{D}}_m^{n}(-\tau)\hat{I}_{\ell}(0) \rangle\nonumber \\
& = 2\int_{-\tau}^{\tau} dt \Big( \langle\hat{I}_B(t)\hat{I}_{\ell}(0)\rangle - \langle\hat{I}_{m}(t)\hat{I}_{\ell}(0)\rangle \Big).
\end{align}
By Eq.~\ref{mixing}, from Eq.~\ref{Dmt} and using the form of $M(\hat{Q},\hat{P})$ in Eq.~\ref{notation2}, we have
\begin{align}
&\lim_{\tau\rightarrow \infty} \int_{-\tau}^{\tau} dt \langle\hat{I}_B(t)\hat{I}_{\ell}(0)\rangle = \lim_{\tau\rightarrow \infty} \int_{-\tau}^{\tau} dt \langle\hat{I}_{m}(t)\hat{I}_{\ell}(0)\rangle \nonumber \\
&\Rightarrow M(\hat{I}_B,\hat{I}_{\ell}) = M(\hat{I}_{m},\hat{I}_{\ell}),
\end{align}

Note that $\hat{I}_{m}$ and $\hat{I}_{\ell}$ are two arbitrary local currents in the system and may be far apart also (for example, $\hat{I}_1$ and $\hat{I}_{L-1}$). So, this rather surprising result tells us that, in the steady state of the open system, the time integrated correlations between any  local current in the system and current from the bath is same as that between any two local currents in the system.
Similar expressions can be derived for energy current and energy current-particle current correlations. Using this and Eq.~\ref{simpf_transp_coeff_local}, we have
\begin{align}
&\left(
\begin{array}{c}
\langle \hat{J}_S \rangle_{NESS}\\
\langle \hat{I}_S \rangle_{NESS}\\
\end{array}
\right)=
(L-1)
\left(
\begin{array}{c}
\langle \hat{J}_{\ell} \rangle_{NESS}\\
\langle \hat{I}_{\ell} \rangle_{NESS}\\
\end{array}
\right) \nonumber\\
& = 
-(L-1)\left(
\begin{array}{cc}
M(\hat{J}_{m},\hat{J}_{\ell}) & M(\hat{I}_{m},\hat{J}_{\ell}) \\
M(\hat{J}_{m},\hat{I}_{\ell}) & M(\hat{I}_{m},\hat{I}_{\ell})\\
\end{array}
\right)
\left(
\begin{array}{c}
\Delta \beta\\
\Delta (-\mu \beta)\\
\end{array}
\right).\label{simpf_transp_coeff1b}
\end{align} 
This is the OCFDR in terms of correlations of local currents inside the system. This very non-trivial result shows that integrated current correlations between any two local currents inside the system is same in the steady state, and gives a transport coefficient. So, even if we look at time integrated correlations between $\hat{I}_1$ and  $\hat{I}_{L-1}$, which are separated by a distance of the order of system length, and even if the system length is large, in the steady state, this correlation is not zero, but is equal to conductance. The thermal steady state of short-range open systems thus harbours long-range correlations.

Finally, summing over $m$ and $\ell$ in Eq.~\ref{simpf_transp_coeff1b} and dividing by $(L-1)^2$, we have the more `expected' form of the result
\begin{align}
&\left(
\begin{array}{c}
\langle \hat{J}_S \rangle_{NESS}\\
\langle \hat{I}_S \rangle_{NESS}\\
\end{array}
\right)\nonumber \\
&=\frac{-1}{L-1}\left(
\begin{array}{cc}
M(\hat{J}_S,\hat{J}_S) & M(\hat{I}_S,\hat{J}_S) \\
M(\hat{J}_S,\hat{I}_S) & M(\hat{I}_S,\hat{I}_S)\\
\end{array}
\right)\left(
\begin{array}{c}
\Delta \beta\\
\Delta (-\mu \beta)\\
\end{array}
\right).
\label{simpf_transp_coeff2}
\end{align}
This is the OCFDR in terms of fluctuations of total currents in the system. This is the form of the OCFDR that would be expected as a naive generalization from the isolated system Green-Kubo formula. It looks very similar to the isolated system Green-Kubo formula. But, there are two important differences. First, it involves equilibrium current fluctuations in presence of the baths. Second, the baths must have infinite degrees of freedom, but the system can be finite. These relations are thus well-defined for small and mesoscopic systems also, unlike those obtained from the isolated system Green-Kubo formula.

Since $M(\hat{Q},\hat{P})=M(\hat{P},\hat{Q})$,  in Eqs.~\ref{simpf_transp_coeff1}, \ref{simpf_transp_coeff1b}, \ref{simpf_transp_coeff2} the Onsager relation $L_{12}=L_{21}$ is satisfied. Thus Onsager relation is not satisfied if the ${\hat{\mathcal{H}}}_S$ is long ranged. So the naive result in Eq.~\ref{simpf_transp_coeff2} does not hold for long ranged systems. But, Eq.~\ref{trasnp_coeff} holds for all cases. For  short-ranged systems, we find that, fluctuations of any current, whether it is the current from the baths, the local currents in the system or the total current in the system, give a transport coefficient upto some system size scaling factors. This is in stark contrast with the standard isolated system Green-Kubo formula, which involves only fluctuations of the total currents in the system.

The Eqs.~\ref{simpf_transp_coeff1b},~\ref{simpf_transp_coeff2} can be understood as generalizations of results in the seminal work of Fisher and Lee Ref.~\cite{FisherLee}. The results for particle conductivity ($L_{22}$) in this section can be obtained from similar fluctuation-dissipation relations in Ref.~\cite{FisherLee} where they are written in frequency space instead of real time. However, our derivation is for a much more general case than that in Fisher and Lee's paper. First, Fisher and Lee's derivation is for non-interacting systems only, i.e, where the whole system+bath Hamiltonian is quadratic. Our derivation shows similar results are valid for interacting systems also, as long as open system thermalization (Eq.~\ref{thermalization}) and the mixing assumption for local currents and densities (Eq.~\ref{mixing}) hold. To our knowledge, this is the first time those results are being generalized to interacting systems. Second, the starting point in Fisher and Lee's calculation requires an external electric field to exist only in the system, even though there is no chemical potential difference between the baths. Though this is rather unphysical, the result of this calculation was shown to match with the NESS calculation using Landauer formula. As recognized in the introduction of Fisher and Lee's paper, this is unsatisfactory. It is not clear why two such different set-ups give the same result.  On the other hand, in our derivation, we started from a system in arbitrary state connected to two thermal baths which have slightly different temperatures and chemical potentials initially. This is the standard set-up to obtain NESS and is much closer to the actual experimental set-ups. Our derivation shows that it is possible obtain Fisher-Lee results by directly doing linear response theory on such set-ups.
Thus, in this section, we have extended previously known OCFDRs to much more general and realistic cases.

In the next section, we will give the connection between the isolated system Green-Kubo formula and the OCFDR in Eq.~\ref{simpf_transp_coeff2}.

\subsection{Crossover between the open-system and the isolated thermodynamic limit}\label{open-closed}
We will be looking at particle conductivity. Similar steps can be followed for OCFDR corresponding to other transport coefficients also.  Let us define the following correlation functions:
\begin{align}
\label{K_m_def}
& \mathcal{K}^O(L,t) = \frac{\beta}{2(L-1)}\int_{-t}^t dt^\prime \langle \hat{I}_S(t^\prime) \hat{I}_S(0) \rangle \nonumber \\
& \mathcal{K}_{p,q}^O(L,t) = \frac{\beta}{2}\int_{-t}^t dt^\prime \langle \hat{I}_p(t^\prime) \hat{I}_q(0) \rangle  \\
& m_2^O(t)=\frac{1}{L-1}\left[\sum_{p,q=2}^{L-1} (p-q)^2 \textrm{Re} \left(\langle \hat{n}_{p}(t) \hat{n}_q(0)\rangle\right) \right] \nonumber \\
& \mathcal{K}^S(L,t) = \frac{\beta}{2(L-1)}\int_{-t}^t dt^\prime \langle\langle \hat{I}_S(t^\prime) \hat{I}_S(0) \rangle\rangle_S \nonumber \\
& m_2^S(t)=\frac{1}{L-1}\left[\sum_{p,q=1}^{L-1} (p-q)^2 \textrm{Re} \left(\langle\langle \hat{n}_{p}(t) \hat{n}_q(0)\rangle\rangle_S\right) \right] \nonumber 
\end{align}
where $\langle\langle ... \rangle \rangle_S$ denotes that the average is taken over the system thermal state $\rho_S = e^{-\beta({\hat{\mathcal{H}}}_S-\mu \hat{N}_S)}/Tr(e^{-\beta({\hat{\mathcal{H}}}_S-\mu \hat{N}_S)})$ and the time translation operator  involves only ${\hat{\mathcal{H}}}_S$. Note that $\mathcal{K}^S(L,t)$ is an isolated system quantity calculated with `free boundary conditions' (as opposed to periodic boundary conditions).  On the other hand, in the first three lines, the averages are over $\rho_{eq}^{\hat{\mathcal{H}}}$, and the time translation operator involves the full system+bath Hamiltonian ${\hat{\mathcal{H}}}$. $\textrm{Re} \left(...\right)$ refers to real part.   The particle conductivity given by the standard Green-Kubo formula is 
\begin{align}
\label{sigmaGK1}
\sigma_{GK} = \lim_{t\rightarrow \infty} \left(\lim_{L \rightarrow \infty} \mathcal{K}^S(L,t)\right).
\end{align}
The order of limits is important and cannot be interchanged.
Our open system result, when ${\hat{\mathcal{H}}}_S$ is short-ranged (Eq.~\ref{simpf_transp_coeff2} with $\Delta \beta=0$), says 
\begin{align}
\label{G}
G\equiv &\lim_{\Delta \mu \rightarrow 0} \frac{\langle I_S \rangle_{NESS}}{(L-1)\Delta \mu}=\frac{1}{L-1}\lim_{t\rightarrow \infty}  \mathcal{K}^O(L,t), \nonumber \\
& = \lim_{t\rightarrow \infty}  \mathcal{K}_{p,q}^O(L,t)
\end{align}
where $G$ is the open system particle conductance.
The open system particle conductivity in the thermodynamic limit is defined as
\begin{align}
\label{sigmaO}
\sigma_O = \lim_{L\rightarrow \infty}(L-1)~G=\lim_{L\rightarrow \infty}\left(\lim_{t\rightarrow \infty}  \mathcal{K}^O(L,t)\right).
\end{align}
Again, the order of limits is important and cannot be interchanged.

Our goal here is to relate $\sigma_O$ and $\sigma_{GK}$. 
To this end, we note that, using Eq.~\ref{continuity}, the following standard result can be shown
\begin{align}
\label{isolated_thermodynamic}
&\lim_{L\rightarrow \infty}\mathcal{K}^S(L,t)= \lim_{L\rightarrow \infty} \frac{\beta}{2} \frac{d}{dt} m_2^S(t).
\end{align}
Using exactly same steps, but for the open system at finite system size, we find that
\begin{align}
\label{main_eqn}
&\mathcal{K}^O(L,t)=\frac{\beta}{2}\frac{d}{dt} m_2^O(t) \nonumber \\
&+\frac{1}{L-1}\sum_{p=1}^{L-1}(2p-1)\left[\mathcal{K}_{p,1}^O(L,t) + \mathcal{K}_{L-p,L-1}^O(L,t)\right] \nonumber\\
&-(L-1) \mathcal{K}_{1,L-1}^O(L,t).
\end{align}
Here, along with the spread of density correlations, we get some boundary terms. Eq.~\ref{main_eqn} is the main result for all further discussions. 

To check the consistency of our calculations, let us first check the long time limit of Eq.~\ref{main_eqn} at finite system size. With $t\rightarrow \infty$, $m_2^O(t)$ reaches a steady state. So the contribution from its derivative is zero. From the second line of Eq.~\ref{G}, we see that each of the boundary terms is proportional to $G$. So, from Eq.~\ref{main_eqn}, we see,
\begin{align}
\label{consistency}
&\lim_{t\rightarrow \infty} \mathcal{K}^O(L,t) = -(L-1)G+\frac{2G}{L-1}\sum_{p=1}^{L-1}(2p-1) \nonumber \\
&=-(L-1)G+\frac{2G}{L-1} (L-1)^2 = (L-1)G,
\end{align}
which is the same as the first line of Eq.~\ref{G}.

Now let us ask what is what happens if the thermodynamic limit of $\mathcal{K}^O(L,t)$ is taken at a fixed $t$. For this, we will require the recently proved \emph{finite temperature Lieb-Robinson bound} \cite{finite_tempLRB}. The main result of the proof, in our context, can be stated plainly as follows. For a system which is short-ranged (in the sense of Eq.~\ref{continuity}), let $\hat{O}_p$ and $\hat{O}_q$ be two local operators with supports at $p$ and $q$ respectively.  Given inverse temperature $\beta$ and a time $t$, there exists a distance $\mid p-q \mid=\mathfrak{L}(\beta,t)$ beyond which the $\langle \hat{O}_p(t)\hat{O}_q(0)\rangle$ decreases exponentially with $\mid p-q \mid$, i.e, 
\begin{align}
\langle \hat{O}_p(t)\hat{O}_q(0)\rangle\sim e^{-\mid p-q\mid},~~\forall~\mid p-q\mid>\mathfrak{L}(\beta,t). 
\end{align}
If $L$ is taken to infinity keeping $t$ finite, $m_2^O(t)$ (see Eq.~\ref{main_eqn}) will not reach its steady state value and will give a major contribution. The contribution of the short-ranged correlations in the boundary terms (for example, $\mathcal{K}_{1,1}^O(L,t)$) is suppressed by the factor of $1/(L-1)$ in front. This factor is not there for the terms involving long-ranged correlations, i.e, terms of the form $\mathcal{K}_{p,q}^O(L,t)$), where $\mid p-q \mid\sim L$. But, as $L$ is increased beyond $\mathfrak{L}(\beta,t)$, these are going to be exponentially suppressed. This means that, with $L\rightarrow \infty$ at finite $t$, the boundary terms will go to zero.  So, we find,
\begin{align}
\lim_{L \rightarrow \infty} \mathcal{K}^O(L,t) = \lim_{L\rightarrow \infty} \frac{\beta}{2} \frac{d}{dt} m_2^O(t).
\end{align}
Looking at the definition of $m_2^O(t)$ in Eq.~\ref{K_m_def}, we see that, again, by finite-temperature Lieb-Robinson bound, only terms where $p$ and $q$ are separated by a finite distance $\delta\ll \mathfrak{L}(\beta,t)$ will have substantial contribution. Let us look at terms where $p-q=\delta$, i.e,
\begin{align}
\label{local_m2}
\frac{1}{L-1}\left[\sum_{q=2}^{L-1} \delta^2 \textrm{Re} \left(\langle \hat{n}_{q+\delta}(t) \hat{n}_q(0)\rangle\right) \right].
\end{align}
Once again, by finite-temperature Lieb-Robinson bound, terms where $q$ in above summation satisfies $\mathfrak{L}(\beta,t)\ll q \ll L-\mathfrak{L}(\beta,t)$ have exponentially small contribution from the baths. We will call these the bulk terms. Thus, for the bulk terms, $\langle \hat{n}_{q+\delta}(t) \hat{n}_q(0)\rangle\simeq \langle\langle \hat{n}_{q+\delta}(t) \hat{n}_q(0)\rangle\rangle_S$. The remaining $\sim 2\mathfrak{L}(\beta,t)$ terms are affected by baths. But, $\mathfrak{L}(\beta,t)$ does not scale with system-size and hence, due to the factor in front, contribution from these terms is suppressed as $1/L$ as $L\rightarrow \infty$. On the other hand, the number of bulk terms scales as $L$, thereby cancelling the $1/L$ factor in front. So, the major contribution comes from the bulk, giving us
\begin{align}
\lim_{L\rightarrow \infty} m_2^O(t)=\lim_{L\rightarrow \infty} m_2^S(t).
\end{align} 
Thus, if the thermodynamic limit is taken at a finite time, we get,
\begin{align}
\label{crossover}
\lim_{L \rightarrow \infty} \mathcal{K}^O(L,t) = \lim_{L\rightarrow \infty} \frac{\beta}{2} \frac{d}{dt} m_2^S(t)=  \lim_{L \rightarrow \infty} \mathcal{K}^S(L,t),
\end{align}
which implies, 
\begin{align}
\label{sigmaGK2}
\sigma_{GK}=\lim_{t\rightarrow \infty}\left(\lim_{L \rightarrow \infty} \mathcal{K}^S(L,t)\right)=\lim_{t\rightarrow \infty}\left(\lim_{L \rightarrow \infty}\mathcal{K}^O(L,t)\right).
\end{align}
Thus, from Eq.~\ref{sigmaO} and ~\ref{sigmaGK2}, we have analytically shown that the $\sigma_{GK}$ and $\sigma_{O}$ are just related by a change in order of the limits taken of the same open system quantity $\mathcal{K}^O(L,t)$. To our knowledge, this is the first time this is being rigorously shown. Moreover, Eqs.~\ref{consistency} and ~\ref{crossover}, we see that $\mathcal{K}^O(L,t)$ shows a crossover from open system behavior to isolated thermodynamic limit behavior with increase in $L$ for fixed $t$, and a crossover from isolated thermodynamic limit behavior to open system behaviour with increase in $t$ for fixed $L$. Note that there is no assumption of weak system-bath coupling.

\subsection{Consequences of the crossover}\label{consequences}
The crossover discussed above has important consequences for the standard methods of classifying transport behavior in open and in isolated systems. Form Eqs.~\ref{sigmaGK1} and \ref{isolated_thermodynamic}, we see that 
\begin{align}
\label{sigma_GK_m2}
\sigma_{GK}= \lim_{L\rightarrow \infty} \frac{\beta}{2} \frac{d}{dt} m_2^S(t).
\end{align}
This standard result is used to classify transport behavior of the isolated system in the thermodynamic limit via time scaling of $m_2^{S}(t)$. Let $m_2^{S}(t)\sim t^{\tilde{\beta}}$. For normal diffusive transport, $\sigma_{GK}$ is finite and $\tilde{\beta}=1$. For $1<\tilde{\beta}<2$, transport is super-diffusive. For $\tilde{\beta}=2$, the transport is ballistic. In both this cases, $\sigma_{GK}$ diverges. For $0<\tilde{\beta}<1$, transport is sub-diffusive and for a localized system $\tilde{\beta}=0$. In both this cases, $\sigma_{GK}=0$. 

On the other hand, scaling of $G$ with $L$ is used to classify open system transport behavior. Let $G \sim L^{-\tilde{\alpha}}$. For normal diffusive transport, $\sigma_O$ is finite and $\tilde{\alpha}=1$. For ballistic transport, $\tilde{\alpha}=0$. For $0<\tilde{\alpha}<1$, transport is super-diffusive. For ballistic and super-diffusive transport, $\sigma_O$ diverges. For $\tilde{\alpha}>1$, transport is sub-diffusive. For a localized system,   $G \sim e^{-L}$. In these two cases, $\sigma_O=0$.  

Thus, our results (Eq.~\ref{G} and ~\ref{crossover}) show that  the open system and the isolated system classifications of transport behvior correspond to the behavior of $\mathcal{K}^O(L,t)$ in different time and length scales. In general, the thermodynamic limit and the long time limit may not commute, leading to different open and isolated system behaviors.

A further interesting insight from above calculations is that $\mathcal{K}_{p,q}^O(L,t)$, $\mid p-q \mid \sim L$, which is proportional to the integrated long-range current-current correlation, must also have clear signatures of both time scaling of $m_2^S(t)$ and system size scaling of conductance $G$. For a given system size, up to some time, this quantity is exponentially small. This time, intuitively, should depend on the time scaling of spread of density correlations $m_2^S(t)$.   At long time, $\mathcal{K}_{p,q}^O(L,t)$ tends to $G$. 

Our treatment can also be taken as an alternate `derivation' of the isolated system Green-Kubo formula. This suggests that the isolated system Green-Kubo formula may not give a transport coefficient if the system Hamiltonian has long range terms. Also, the analogous derivation for thermal currents gives a `derivation' of the isolated system thermal conductivity formula without any assumption of local equilibrium \cite{Mori1958} or `gravitational field' \cite{Luttinger1964}.

This brings us to the end of the analytical part of the paper. At this point, it is worth re-iterating the assumptions made in deriving all the results,

a) Ths full system+bath set-up, i.e, ${\hat{\mathcal{H}}}_S+ {\hat{\mathcal{H}}}_{SB}+{\hat{\mathcal{H}}}_{B}$, is time translation and time reversal invariant.

b) Open system thermalization (Eq.~\ref{thermalization}).

c) ${\hat{\mathcal{H}}}_S$ is short-ranged.

d) Mixing assumption for local currents and densities (Eq.~\ref{mixing}).

No other assumptions have been made. The forms of ${\hat{\mathcal{H}}}_S$, ${\hat{\mathcal{H}}}_{SB}$ and ${\hat{\mathcal{H}}}_{B}$ are arbitrary. The high temperature limit of the results give the classical results. In the following, we will apply our theory to a numerically tractable but non-trivial example.

\begin{figure}
\includegraphics[width=\columnwidth]{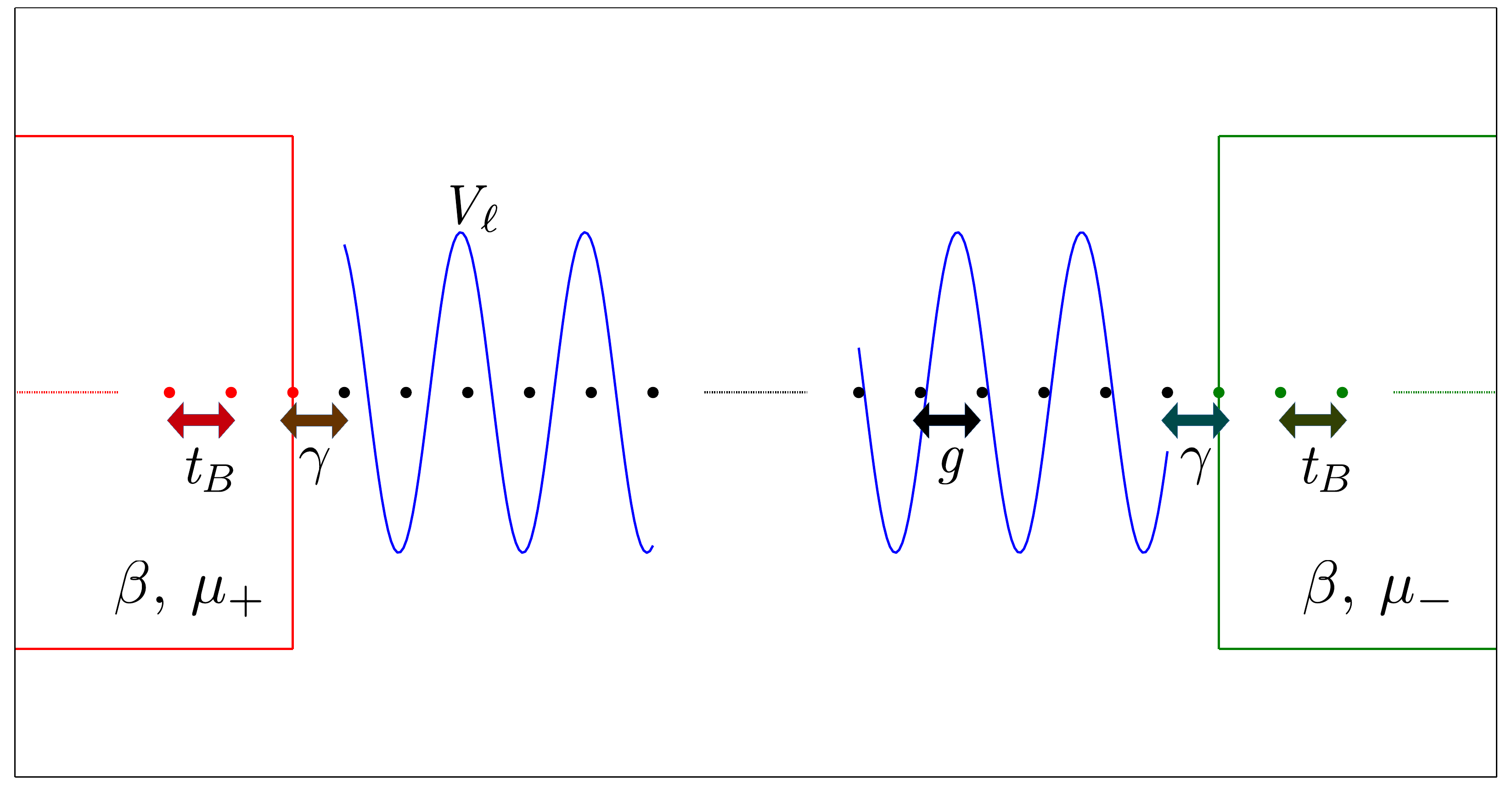}
\caption{(color online)  Our non-trivial but numerically tractable example. The system consists of the critical AAH model. $V_\ell$ is quasiperiodic potential of the critical AAH model, $V_\ell = 2 \cos(2\pi b \ell+\phi)$, where $b$ is an irrational number. The hopping parameter in the system is $g$ which is set to $g=1$. The baths are attached at the first and the last system sites. The baths are modelled by semi-infinite ordered non-interacting tight-binding chains with hopping parameter $t_B$. The hopping between the system and the baths are given by $\gamma$. The initial state of the set-up is taken to be of the same form as the in Fig.~\ref{fig:set_up}, but with no temperature bias, i.e, $\beta_+=\beta_-=\beta$. In the numerics, $b$ is chosen as the golden mean, $b=(\sqrt{5}-1)/2$, and we only look at the particle transport.}
\label{fig:AAH_set_up}
\end{figure}

\begin{figure*}
\includegraphics[width=\linewidth]{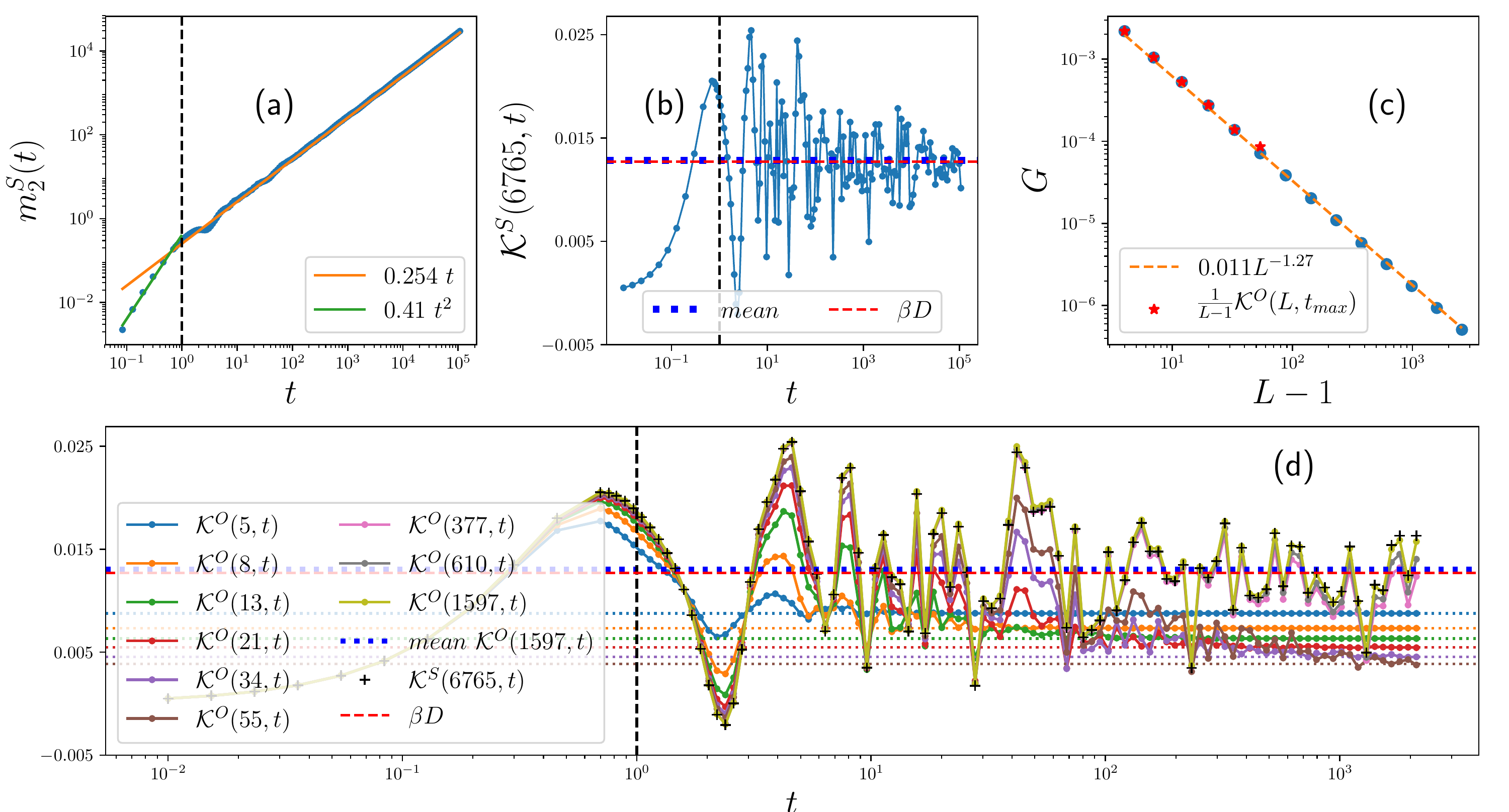}
\caption{(color online) \textbf{(a)} Plot of $m_2^{S}(t)$ for the critical AAH model. At very short time, $t<1$ (vertical black dashed line), $m_2^{S}(t)$ scales ballistically, i.e, $m_2^{S}(t)\propto t^2$. For $t>1$,  $m_2^{S}(t)$ scales diffusively, i.e, $m_2^{S}(t)= 2Dt$. From the fit, $D=0.254/2=0.127$. System-size, $L=6765$. \textbf{(b)} Plot of $\mathcal{K}^S(L,t)$ for $L=6765$. $\mathcal{K}^S(L,t)$ rises initially and then saturates showing fluctuations about a mean value. The mean value is $\beta D$, akin to a diffusive system. The vertical line gives the time $t=1$ after which diffusive scaling of $m_2^{S}(t)$ starts. The mean is calculated from the data points to the right of the vertical line. \textbf{(c)}  The sub-diffusive scaling of particle conductance $G$ with system size calculated using Eq.~\ref{Landauer}. Also shown are values of $\mathcal{K}^O(L,t_{max})/(L-1)$ for $L\leq 55$. Here $t_{max}$ is the final time point in (d). \textbf{(d)} Plots of $\mathcal{K}^O(L,t)$ for various system sizes. In the time range considered, for $L\leq 55$, the steady state is reached. The small-dotted lines show the corresponding values of $(L-1)~G$ from (c). For much larger system sizes, in this time range, $\mathcal{K}^O(L,t)$ converges to $\mathcal{K}^S(6765,t)$ (the black `$+$' symbols). The vertical dashed line corresponds to $t=1$, the same as in (a). The mean of data points for $t>1$ for $L=1597$ is shown with blue squares. It agrees quite well with $\beta D$ (the red dashed horizontal line). Parameters: Bath length $L_B=3500$, $\beta=0.1$, $\mu=1$, $\gamma=1.5$, $t_B=1.5$, $t_{max}=2123$. The unit of time is the system hopping parameter which has been set to $1$.}
\label{fig:AAH_all}
\end{figure*}

\subsection{A non-trivial example: critical AAH model}\label{AAH}
Our theory is especially important for cases where open system classification and isolated system classification of transport behaviors give different results. Now we explore in detail the critical AAH model which, as discussed in the introduction, is one such example. The critical AAH model \cite{aa1,harper} Hamiltonian is given by
\begin{equation}
\label{H_S_AAH}
\hat{\mathcal{H}}_S~=~\sum_{\ell=1}^{L-1}(\hat{c}_\ell^{\dagger} \hat{c}_{\ell +1}+h.c)+\sum_{\ell=1}^L 2 \cos(2\pi b \ell+\phi) \hat{c}_\ell^{\dagger} \hat{c}_\ell 
\end{equation}
where $b$ is an irrational number, $\phi$ is an arbitrary phase, and $\hat{c}_\ell $ is the fermionic annihilation operator at site $\ell$. The eigenstates of this model are neither totally delocalized nor localized, but are `critical' \cite{pandit83}. This holds true for any choice of irrational number $b$ and phase $\phi$. This model and its various generalizations have been of recent interest in both theoretical \cite{oc_mismatch3,vkv,Yang2017,david2017,Naldesi2016,AAH2017,AAH4,AAH5,
AAH6,AAH7,AAH8,AAH9,AAH10,AAH11,AAH12,AAH13,AAH14,AAH15,AAH16,AAH17,AAH18,AAH19} and experimental \cite{expt0,expt1,expt2,expt3,Zilberberg2013,Zilberberg2013_2,expt4,expt5,expt6} fronts. It can also be derived from a 2D system under a magnetic field (quantum-Hall like set-up) {\cite{Hofstader,Zilberberg2012,Zilberberg2013}.

The transport properties of this model in both open and isolated set-ups have been thoroughly studied very recently in Ref.~\cite{oc_mismatch3,vkv}. It has been shown that $m_2^{S}(t)\sim t$ like a diffusive system, but the scaling of NESS conductance with system size is sub-diffusive. (Actually, the isolated system is not strictly diffusive, but have some hints of super-diffusive behavior, see Appendix~\ref{AAH_transport}.) Thus, the open and  the isolated system classifications of transport are inconsistent in this model.  As discussed in Ref.~\cite{oc_mismatch3}, the reason for this is that the single-particle eigenfunctions of the critical AAH model has very different behavior in the bulk and in the edges. We want to investigate what this entails for the equilibrium current correlations of the open system. By our discussion above, this is an explicit example where the thermodynamic limit and the long-time limit do not commute (see Eqs.~\ref{sigmaO},~\ref{sigmaGK2}), thereby providing a non-trivial test-bed for our theory.

It has also been shown that the sub-diffusive scaling exponent of particle conductance changes depending on the choice of system sizes (though always remaining sub-diffusive) \cite{oc_mismatch3,vkv}. For our exact numerical calculations below, we will choose $b$ as the golden mean $(\sqrt{5}-1)/2$ and take the system sizes equal to Fibonacci numbers. All our results will be averaged over $\phi$ so that translational invariance is restored. 

To calculate the open system quantities, we will choose the following model of for the baths and the system-bath couplings,
\begin{align}
\label{baths}
& {\hat{\mathcal{H}}}_{B_1}= t_B (\sum_{s=-\infty}^0 \hat{b}_s^{(1)\dagger}\hat{b}_{s+1}^{(1)}+h.c.), \\
& {\hat{\mathcal{H}}}_{B_2}= t_B (\sum_{s=L+1}^\infty \hat{b}_s^{(2)\dagger}\hat{b}_{s+1}^{(2)}+h.c.), \nonumber \\
& {\hat{\mathcal{H}}}_{SB_1} = \gamma (\hat{c}_1^{\dagger}\hat{b}_{0}^{(1)}+ h.c.),  \hspace{5pt} {\hat{\mathcal{H}}}_{SB_2}=\gamma(\hat{c}_L^{\dagger}\hat{b}_{L+1}^{(2)}+ h.c.). \nonumber
\end{align}
Thus, left bath consists of sites from $-\infty$ to $0$ (with fermionic annihilation operators $\hat{b}_{s}^{(1)}$), the sites from $1$ to $L$ is our system (with fermionic annihilation operators $\hat{c}_{\ell}$, see Eq.~\ref{H_S_AAH}), while the sites from $L+1$ to $\infty$ is our right bath (with fermionic annihilation operators $\hat{b}_{s}^{(2)}$). The two baths have same hopping parameter $t_B$.  Thus the baths are modelled by semi-infinite ordered non-interacting tight-binding chains, the spectral functions of which are well approximated by continuous functions. The hopping parameter $t_B$ is chosen such that the bandwidths of the baths are larger than that of the system. Hence, the conditions for showing open system thermalization are satisfied \cite{dharsen2006,dharroy2006}.  The system-bath coupling to each bath is the hopping from the system to the bath, given by the parameter $\gamma$. See Fig.~\ref{fig:AAH_set_up} for a schematic of the set-up. 
For this set-up, the particle current operators are given by
\begin{align}
\label{current_op}
\hat{I}_{p} = i(\hat{c}_p^{\dagger} \hat{c}_{p+1} - \hat{c}_{p+1}^{\dagger} \hat{c}_{p}), \ \hat{I}_S = \sum_{p=1}^{L-1} \hat{I}_{p}.
\end{align}
We can calculate $G$ exactly using the formula,
\begin{align}
\label{Landauer}
G =  \int \frac{d\omega}{2\pi} T(\omega) \mathfrak{n}^2(\omega) e^{\beta(\omega-\mu)}
\end{align}
where $\mathfrak{n}(\omega) = [e^{\beta(\omega-\mu)}+ 1]^{-1}$ is the Fermi distribution and $T(\omega)$ is the transmission function which can be exactly calculated as given in Appendix.~\ref{transmission}. 
To calculate $\mathcal{K}^S(L,t)$, $m_2^S(t)$, we use exact diagonalization of ${\hat{\mathcal{H}}}_S$. We obtain $\mathcal{K}^O(L,t)$, $\mathcal{K}_{p,q}^O(L,t)$ by exact diagonalization of full system+bath Hamiltonian ${\hat{\mathcal{H}}}$ by considering finite but large baths, and looking at times before the finite size effects of the bath become significant.

For completeness, let us first check the dramatic difference between open system and isolated system classifications of transport behavior of the model. The scaling of $m_2^{S}(t)$ is shown in Fig.~\ref{fig:AAH_all}(a). At extremely small time $m_2^{S}(t)$ shows ballistic scaling $m_2^{S}(t)\sim t^2$. At longer times, $m_2^{S}(t)$ shows an almost perfect diffusive scaling
\begin{align}
\label{diffusive_scaling}
m_2^{S}(t)=2Dt,~~D=0.127
\end{align}
Here $D$ is the diffusion constant which is extracted from the fit. The crossover from ballistic to diffusive scaling occurs at $t\sim 1$ (the vertical dashed line in Fig.~\ref{fig:AAH_all}(a)).

From Eq.~\ref{sigma_GK_m2}, we see that 
\begin{align}
\sigma_{GK}=\beta D.
\end{align}
From the definition of $\sigma_{GK}$ (Eq.~\ref{sigmaGK1}) we expect that $\mathcal{K}^S(L,t)$ will saturate to this value for large enough systems and at long enough times. This is shown in Fig.~\ref{fig:AAH_all}(b). During the time which corresponds to the initial ballistic spread of  $m_2^{S}(t)$, $\mathcal{K}^S(L,t)$ rises. After that, i.e, for $t>1$, $\mathcal{K}^S(L,t)$ saturates showing fluctuations about a mean value. The fluctuations decrease with time. The mean of data points for $t>1$ is almost exactly given by $\beta D$. Thus, the diffusive-like behavior in terms of the isolated system classification is established (see Appendix~\ref{AAH_transport}). For both Fig.~\ref{fig:AAH_all}(a) and Fig.~\ref{fig:AAH_all}(b), the system size is $L=6765$.

In Fig.~\ref{fig:AAH_all}(c), we show the system-size scaling of open system particle conductance $G$, as calculated using Eq.~\ref{Landauer}. $G$ shows an almost perfect sub-diffusive scaling
\begin{align}
G\sim L^{-1.27\pm0.01},
\end{align}
as previously shown in Ref.~\cite{oc_mismatch3}.
Thus, the stark difference between the open system and the isolated system classifications of transport in this model is very clear. In the following, let us see what this entails for the equilibrium current fluctuations of the open system.

In Fig.~\ref{fig:AAH_all}(d), we show all plots of $\mathcal{K}^O(L,t)$ for various system sizes. The time range taken is from $0.01$ to $t_{max}=2123$ (in units of the hopping parameter). Up to this time, in the numerics, there was no effect of finite bath size. In this time range, $\mathcal{K}^O(L,t)$ reaches the steady state value for $L\leq 55$. The steady state value is quite precisely given by $(L-1) G$ as shown by the dotted lines in Fig.~\ref{fig:AAH_all}(d). This is also shown in Fig.~\ref{fig:AAH_all}(c), where $\mathcal{K}^O(L,t_{max})/(L-1)$, for $L \leq 55$, has been plotted on top of the exactly calculated $G$. The sub-diffusive scaling of the steady state  values of $\mathcal{K}^O(L,t_{max})$ (i.e, when $t\rightarrow \infty$ is taken first) is clear. ( The fact that $\mathcal{K}^O(L,t)$ reaches a steady state value given by $(L-1)G$ means that the mixing assumption (Eq.~\ref{mixing}) is valid. We have also explicitly checked this in Appendix.~\ref{check_mixing}.) On the other hand, on increasing system size, in the time range considered (i.e, when $L\rightarrow \infty$ is taken first), $\mathcal{K}^O(L,t)$ converges to $\mathcal{K}^S(L,t)$, which shows fluctuations about the mean value $\beta D$. This is shown by plotting $\mathcal{K}^O(L,t)$ for $L=377,610,1597$ in the same time range and comparing with $\mathcal{K}^S(6765,t)$. The data points for $\mathcal{K}^O(1597,t)$ and $\mathcal{K}^S(6765,t)$ are almost overlapping.  The mean of data points for $t>1$ for $L=1597$ is also shown, and it agrees quite well with $\beta D$. 

This shows that both the diffusive scaling spread of correlations and the sub-diffusive scaling of current are encoded in $\mathcal{K}^O(L,t)$. Indeed, they correspond to taking the thermodynamic limit ($L\rightarrow \infty)$ and  the long time ($t\rightarrow \infty$) of $\mathcal{K}^O(L,t)$ in different orders. Thus, $\sigma_O$ and $\sigma_{GK}$ are indeed related by a change in the order of limits and, as in the present case, the limits may not commute.

\begin{figure}
\includegraphics[width=\columnwidth]{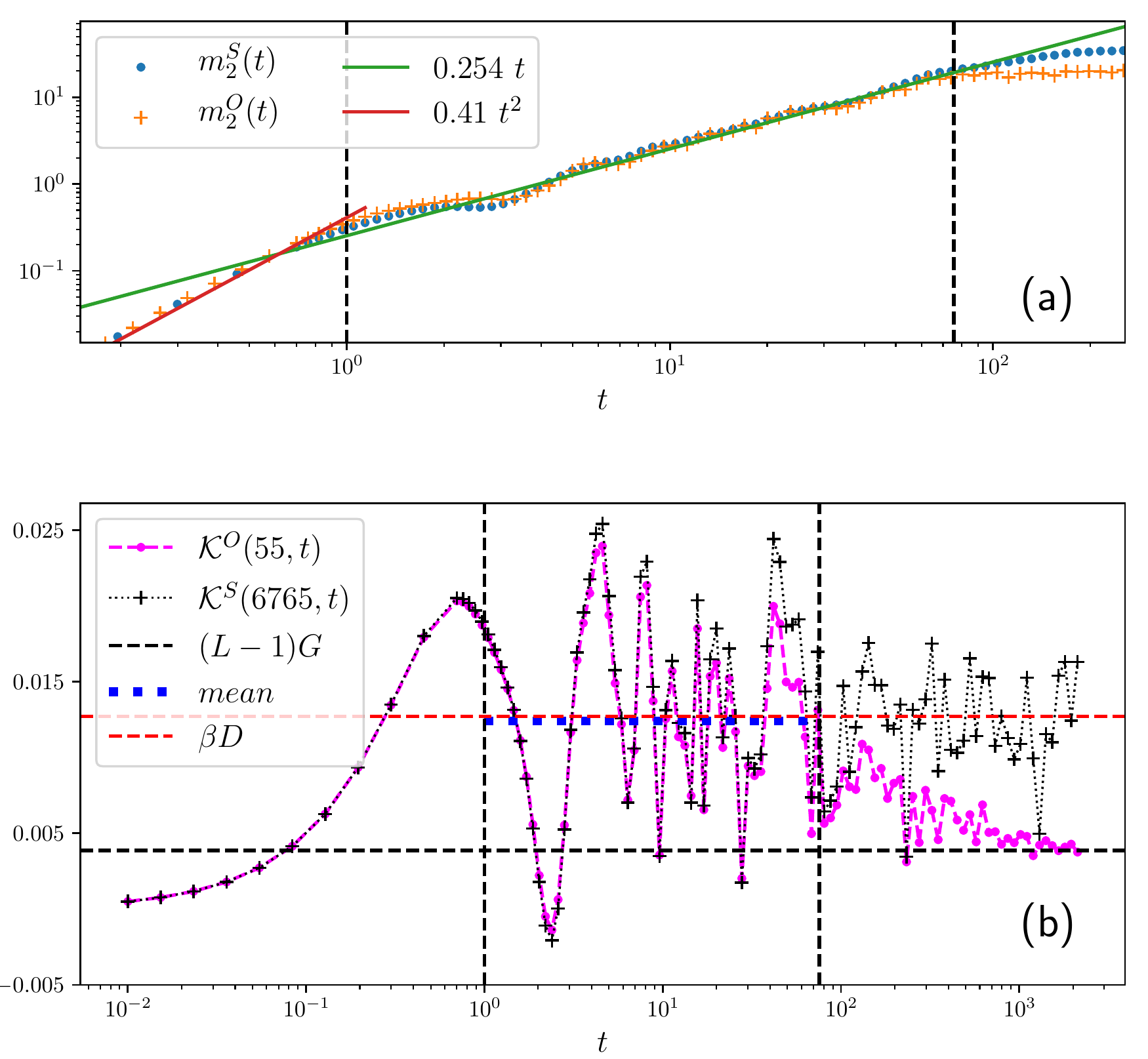}
\caption{(color online) \textbf{(a)} Plot of $m_2^{S}(t)$ and $m_2^{O}(t)$  for system size $L=55$. The vertical dashed lines denote the time range during which the diffusive scaling holds. Beyond this time finite system size effects are seen in $m_2^{S}(t)$, and $m_2^{O}(t)$ reaches a steady value. \textbf{(b)} Plot of $\mathcal{K}^O(55,t)$  which is compared with $\mathcal{K}^S(6765,t)$. The vertical dashed lines denote the same time range as in (a). The mean of data points for $\mathcal{K}^O(55,t)$ in this time range (blue squares) agrees well with $\beta D$. In fact, in this time range, $\mathcal{K}^O(55,t)$ and  $\mathcal{K}^S(6765,t)$ agree well. Beyond this time, $\mathcal{K}^O(55,t)$ decays to its steady state value given by $(L-1) G$. Other parameters are same as in Fig.~\ref{fig:AAH_all}. }
\label{fig:AAH_55}
\end{figure}

As is evident, in the present case, whether the diffusion-like behavior is seen or the sub-diffusive behavior is seen depends on the length and the time scales one is looking at. Let us now look at the time scales in more detail for $L=55$.  Plots of $m_2^{S}(t)$ and $m_2^{O}(t)$ for $L=55$ is given in Fig.~\ref{fig:AAH_55}(a). The first thing to note is that the diffusive scaling starts at $t\sim 1$ which is the same as in Fig.~\ref{fig:AAH_all}(a). Thus, this time scale is independent of system size. The diffusive scaling of $m_2^{S}(t)$ is seen up to some time $t^*$, after which finite system size effects occur. For time less than $t^*$, $m_2^{S}(t)$ and $m_2^{O}(t)$ match. After time $t^*$, both $m_2^{S}(t)$ and $m_2^{O}(t)$ show finite system-size effects, $m_2^{O}(t)$ reaching a steady value.  The time range for diffusive scaling of $m_2^{S}(t)$ is demarcated in  Fig.~\ref{fig:AAH_55}(a) via the vertical dashed lines. It is exactly in this time range that the $\mathcal{K}^O(55,t)$ also shows the diffusive-like behavior. This is shown in Fig.~\ref{fig:AAH_55}(b). In the same time range demarcated by the vertical dashed lines, $\mathcal{K}^O(55,t)$ shows fluctuations about a mean value. The mean of the data points in this time range agrees well with $\beta D$. In fact, $\mathcal{K}^O(55,t)$ and $\mathcal{K}^S(6765,t)$ match well for $t<t^*$. For $t>t^*$, $\mathcal{K}^O(55,t)$ decays to its steady state value which is given by $(L-1) G$. Since $m_2^{S}(t)\propto t$, $t^*$ scales as $\sim L^2$ with system size.

Thus, for a given system size, the open critical AAH model shows signatures of diffusive transport in the integrated equilibrium  total current fluctuations in some time range. This time range corresponds to the time range over which diffusive spread of correlations in the isolated system of same size is seen. This time range grows with system size as $\sim L^2$.  Beyond this time scale, the effect of the baths start to matter, and the integrated total current fluctuations reach a steady state.  The system size scaling of the steady state values of the integrated total current fluctuations is sub-diffusive.

\begin{figure}
\includegraphics[width=\columnwidth]{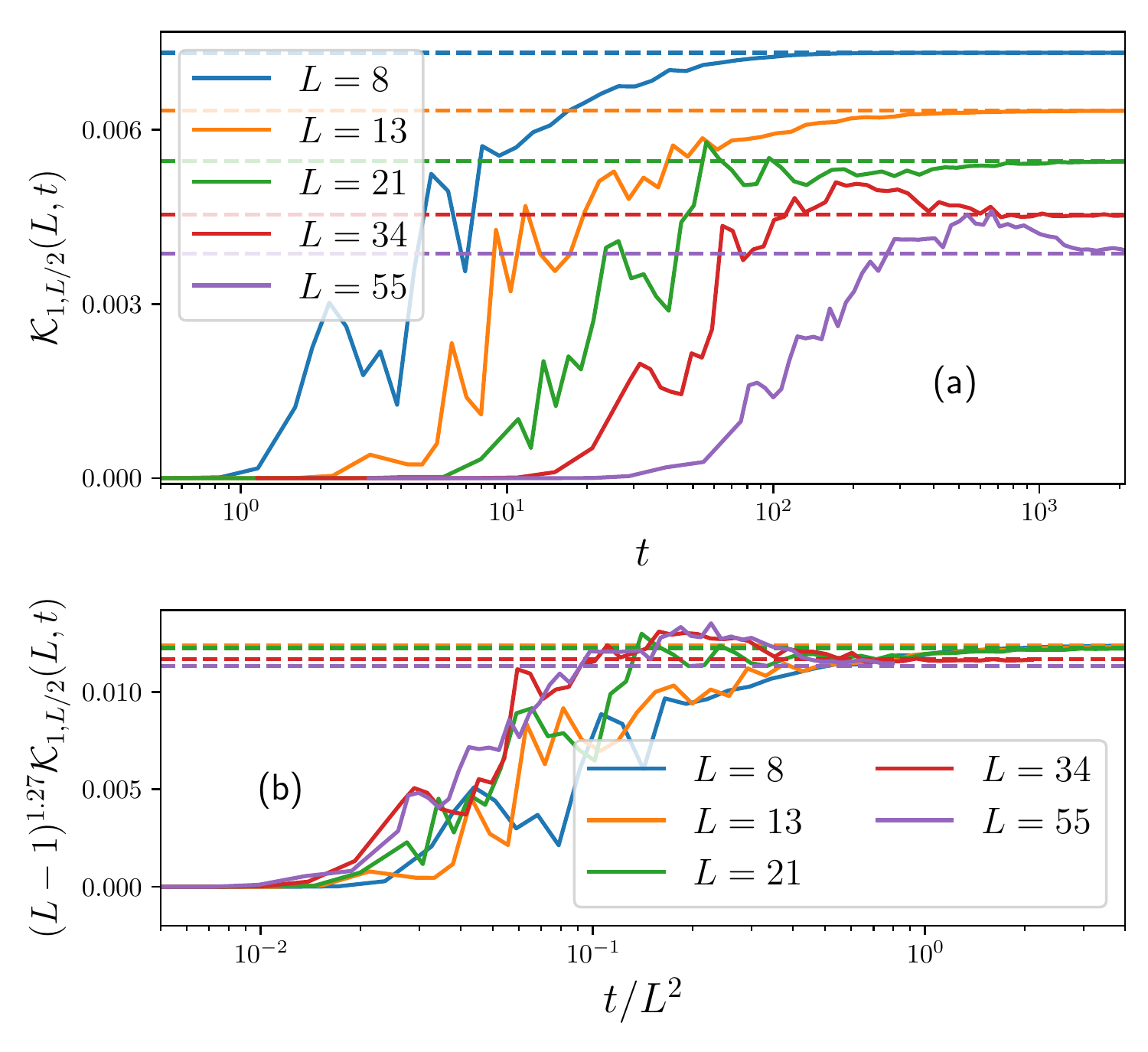}
\caption{(color online) \textbf{(a)} Plots of the integrated long range current correlations $\mathcal{K}_{1,L/2}(L,t)$ with $t$ for different system sizes. The horizontal dashed lines show the corresponding values of conductance $G$ calculated exactly. \textbf{(b)} The scaled plots of $\mathcal{K}_{1,L/2}(L,t)$. To collapse the x-axis, the $t$ needs to be scaled diffusively (consistent with $m_2^S(t)=2Dt$). To collapse the y-axis, $\mathcal{K}_{1,L/2}(L,t)$ needs to scaled `sub-diffusively' (consistent with $G\sim L^{-1.27\pm 0.01}$).    Other parameters are same as in Fig.~\ref{fig:AAH_all}. }
\label{fig:AAH_non_local}
\end{figure}

Finally, let us look at the integrated long range current correlations $\mathcal{K}^O_{1,L/2}(L,t)$. According to our theory, this quantity should also have information about both the diffusive behavior of the isolated system and the sub-diffusive behavior of the open system. Fig.~\ref{fig:AAH_non_local}(a) shows plots of $\mathcal{K}^O_{1,L/2}(L,t)$ with time for various system sizes. As expected from our discussions, $\mathcal{K}^O_{1,L/2}(L,t)$ starts to be substantial only after a finite time. This time grows with system size. It corresponds to the time required for correlations to spread from site $1$ to site $L/2$ inside the system. Hence, this time is expected to scale diffusively with system size, consistent with scaling of $m_2^S(t)$. On the other hand, in long time limit, $\mathcal{K}^O_{1,L/2}(L,t)$ reaches a steady state value precisely given by the corresponding conductance $G$, as expected from our theory. So the steady state value should scale sub-diffusively with system-size. As a result, $\mathcal{K}^O_{1,L/2}(L,t)\sim L^{-1.27} f(t/L^2)$, and we expect a data collapse. The approximate data collapse of the scaled plot is shown in Fig.~\ref{fig:AAH_non_local}(b). The collapse is not so good in the growing part of $\mathcal{K}^O_{1,L/2}(L,t)$ due to fluctuations. The fluctuations in the growing part of $\mathcal{K}^O_{1,L/2}(L,t)$ seems to decrease with system size, but larger system sizes are required for conclusive evidence regarding this. Nevertheless, it is clear that the long range current correlations of the open critical AAH model also shows signatures of both diffusion and sub-diffusion.

Two points are worth mentioning before we close this section. The first is that the signatures of diffusive transport in the current fluctuations of the isolated system are presumably a transient behavior. As shown in Ref.~\cite{oc_mismatch3}, the time scaling of higher moments show a diffusive to super-diffusive crossover. The time for this crossover is smaller for higher moments. It is presumable that $m_2^{S}(t)$ will also show super-diffusive scaling at extremely long times. By Eq.~\ref{isolated_thermodynamic}, this should also show up in $\mathcal{K}^S(L,t)$. But, as given in Ref.~\cite{oc_mismatch3}, the time required to observe this in numerics is estimated to be $>10^{10}$ (in steps of the system hopping parameter which is set to 1). The system-size required to observe this in numerics without having finite-size effects is estimated to be $\sim 10^7$. This is beyond our current computational ability. Nevertheless, as we have shown above, for finite-size open critical AAH model, the transient `diffusive-like' behavior of the isolated system and the sub-diffusive behavior of the open system will both be captured in the time and the system-size dependence of current fluctuations of the system.

The second is that, in this section, we have considered a set-up where the system is a critical AAH model while the baths are semi-infinite ordered nearest neighbour tight-binding chains.  The reader may be curious about what would happen if the baths were also taken as the critical AAH model. In that case, the theory cannot be applied as the set-up will not show open system thermalization (Eq.~\ref{thermalization}). This is because, critical AAH model has a fractal spectrum, which cannot be approximated by a continuous function even if the system size is infinite. Thus the bath spectral functions would not be continuous functions. This violates one of the required conditions for non-interacting systems to show open system thermalization (see discussion following Eq.~\ref{thermalization} \cite{dharroy2006,dharsen2006}).

\subsection{Summary and outlook}\label{conclusions} 

In this paper, we have obtained several important and fundamental results in non-equilibrium statistical physics. In conclusion, we first give all the rigorous analytical results point-by-point, clearly mentioning the assumptions required for each.
\begin{itemize}
\item 
\emph{Assumption 1}: Time-translation and time-reversal invariance of the system, bath and system-bath coupling Hamiltonians.

\emph{Assumption 2}: Open system thermalization (Eq.~\ref{thermalization}).

 \emph{Result 1}: Rigorously showing that, in linear response regime, expectation values of system operators reach a unique NESS value given by Eq.~\ref{O_NESS}, irrespective of the initial state of the system.

\emph{Result 2}: Obtaining the OCFDR for total system currents (Eq.~\ref{trasnp_coeff}), which is valid for interacting and non-interacting, short-ranged and long-ranged systems. At this level, the Onsager relation is not valid in general.

\item \emph{Assumption 3:} Short-ranged systems (Eq.~\ref{continuity}).

\emph{Result 1:} Showing the equivalence between OCFDR for currents from the baths and OCFDR for total system currents (Eq.~\ref{simpf_transp_coeff1}), and recovery of Onsager relations.

\item \emph{Assumption 4:} Mixing assumption for local currents and densities (Eq.~\ref{mixing}):

\emph{Result 1}: Showing that the time integrated current-current correlation between \emph{any two local currents} of the open system in equilibrium is the \emph{same} and is proportional to the corresponding transport coefficient (Eq.~\ref{simpf_transp_coeff1b}). The OCFDRs in Eqs.~\ref{simpf_transp_coeff1},~\ref{simpf_transp_coeff1b},~\ref{simpf_transp_coeff2} are may be expected from previous investigations for non-interacting systems \cite{FisherLee, shot_noise, Abhishek2}. We have rigorously extended them to interacting systems via a quite robust derivation.

\emph{Result 2}: Rigorously showing that transport coefficients obtained from the isolated system Green-Kubo formula and from the OCDFR are related by a change in the order of taking the thermodynamic and the long time limits of the integrated total system current correlations of the open system (Eqs.~\ref{sigmaO} and~\ref{sigmaGK2}). This means that the time and system size dependence of equilibrium current correlations of the open system can be used to classify transport behaviors of both the open system and the isolated system.
\end{itemize}

We have then numerically checked the implications of the above result for the critical AAH model, where it has been recently shown that transport is diffusive-like according to isolated system classification but sub-diffusive according to open system classification. We considered the critical AAH model connected to two baths modelled by infinite 1D nearest neighbour non-interacting tight-binding chains. The important results here are:
\begin{itemize}
\item The integrated total current autocorrelation of the open system $\mathcal{K}^O(L,t)$ (see Eq.~\ref{K_m_def}) shows signatures of diffusive-like behavior up to a time scale. This time scale grows as $L^2$, where $L$ is the system length, which is consistent with the diffusive-like behavior. In later times, it reaches a steady value. The steady state value is exactly $L-1$ times the conductance, which can be independently calculated from NEGF. The conductance scales sub-diffusively with system size (Figs.~\ref{fig:AAH_all},~\ref{fig:AAH_55}).
\item The integrated long-range current correlations of the open system $\mathcal{K}^O_{1,L/2}(L,t)$ (see Eq.~\ref{K_m_def}) also shows both diffusive-like and sub-diffusive behaviors. It is zero up to a time scale which again scales as $L^2$, showing the diffusive propagation. Then it rises and finally reaches a steady state. The steady state value is given by conductance, which shows sub-diffusive scaling with system size (Fig.~\ref{fig:AAH_non_local}).
\end{itemize}
To the best of our knowledge, this is the first work where time and system-size dependence of open system equilibrium current correlations are being used to classify both isolated system and open system transport behaviors of a model. We believe that, specifically, the role of long-range current correlations in this respect is crucial to understand and requires further investigations. Our derivations are completely general and works for interacting systems also. However, the example we have worked out numerically is a non-interacting case, though quite non-trivial. This is because similar direct numerical investigation is extremely challenging in presence of interactions. In future works, we will attempt to rise to that challenge and check our results for interacting systems. Another non-trivial direction is to check the theory for long-range non-interacting systems, where according to our theory, the Onsager relation may not be valid for system currents. Investigations in this direction are under progess.

\textit{Acknowledgements: } The author would like to thank Abhishek Dhar, Anupam Kundu, Sumilan Banerjee and Aritra Kundu for extremely useful discussions.

\vspace{45pt}

\section*{Appendix}

\appendix

\section{Open system thermalization and eigenstate thermalization}\label{ETH}
Note the stark contrast between the open system thermalization statement given in Eq.~\ref{thermalization} and the ETH statement \cite{ETH_review}. ETH does not hold for non-interacting systems, while the open system thermalization statement holds. This is because, though the initial state of the system ($\rho_S$) is arbitrary, $\rho_{EIS}^{\hat{\mathcal{H}}}$ (Eq.~\ref{rho_notations}) is a special form of initial state for the whole system+bath set-up. For non-interacting systems, Eq.~\ref{thermalization} does not hold for system operators if the initial state of the whole set-up is not of this special form. It is thus consistent with the fact that ETH, which considers more generic initial states, does not hold for non-interacting systems. However, initial states of the form $\rho_{EIS}^{\hat{\mathcal{H}}}$, being physically motivated, are widely used as the starting point in open system calculations to discuss equilibriation.

\section{$M(\hat{Q},\hat{P})=M(\hat{P},\hat{Q})$}\label{MqpMpq}

This result was stated after Eq.~\ref{notation}, and it was mentioned that it holds when $H$ has time-reversal and time translation invariance. Here we present the proof. Let $\mathcal{T}$ be the time reversal operator. \begin{align}
&\langle \hat{Q}(t)\hat{P}(t^\prime) \rangle = \langle \mathcal{T} \hat{Q}(t)\hat{P}(t^\prime)\mathcal{T}^{-1} \rangle \nonumber \\
&= \langle \hat{P}(-t^\prime)\hat{Q}(-t) \rangle =\langle \hat{P}(-t^\prime+\tau)\hat{Q}(-t+\tau) \rangle
\end{align}
where the last line is the statement of time-translation invariance. The choice of $\tau=t+t^\prime$ gives $\langle \hat{Q}(t)\hat{P}(t^\prime) \rangle = \langle \hat{P}(t)\hat{Q}(t^\prime) \rangle$. With this property, it is obvious that $M(\hat{Q},\hat{P})=M(\hat{P},\hat{Q})$.

\section{The simplification of Eq.~\ref{notation} to Eq.~\ref{notation2}}\label{simpf_Mqp}
Here we give the simplification from the Eq.~\ref{notation} to Eq.~\ref{notation2}. For this simplification, we need the following two results. The first is:
\begin{align}
\label{KMS}
&\langle \hat{Q}(t)\hat{P}(t) \rangle = \frac{Tr(e^{-\beta(\hat{H}-\mu \hat{N})}\hat{Q}(t)\hat{P}(t))}{Z} \nonumber \\
&= \frac{Tr(e^{-\beta \hat{H}}\hat{Q}(t)e^{\beta \hat{H}}e^{-\beta(\hat{H}-\mu \hat{N})}\hat{P}(t))}{Z} \nonumber\\
&= \langle \hat{P}(t)\hat{Q}(t+i\beta) \rangle,
\end{align}
where in the second line we have used the fact that $[\hat{Q},\hat{N}]=0$, which is true for particle and energy current operators.  We also require that the following limit exists
\begin{align}
\lim_{\tau\rightarrow\infty} \int_{-\tau}^\tau dt \langle \hat{Q}(- i\lambda) \hat{P}(t) \rangle = \lim_{\tau\rightarrow\infty} \int_{-\tau}^\tau dt  \langle \hat{Q}(0) \hat{P}(t+i\lambda) \rangle.
\end{align}
For this, it is necessary that
\begin{align}
\label{necessary}
\lim_{t\rightarrow \pm\infty} \langle \hat{Q}(0) \hat{P}(t+i\lambda) \rangle = 0
\end{align}
Now, we can simplify the expression for $M(\hat{Q},\hat{P})$ as the following:
\begin{align}
&\beta M(\hat{Q},\hat{P}) = \int_0^\infty dt \int_0^\beta d\lambda \langle \hat{Q}(- i\lambda) \hat{P}(t) \rangle \nonumber \\
&=\int_0^\infty dt \int_0^\beta d\lambda \langle \hat{P}(t) \hat{Q}( i(\beta-\lambda))  \rangle \hspace{5pt} \textrm{(Using Eq.~\ref{KMS})} \nonumber \\
& = \int_0^\infty dt \int_0^\beta d\lambda \langle \hat{P}(t) \hat{Q}( i\lambda)  \rangle \hspace{20pt} \textrm{($\lambda \rightarrow \beta-\lambda$)} \nonumber \\
& = \int_0^\infty dt \int_0^\beta d\lambda \langle \hat{Q}(- i\lambda) \hat{P}(-t) \rangle \hspace{5pt} \textrm{(Using time-reversal)} \nonumber \\
&=\int_{-\infty}^0 dt \int_0^\beta d\lambda \langle \hat{Q}(- i\lambda) \hat{P}(t) \rangle \hspace{17pt} \textrm{($t\rightarrow -t$)} \nonumber \\
&=\frac{1}{2}\int_{-\infty}^\infty dt \int_0^\beta d\lambda \langle \hat{Q}(- i\lambda) \hat{P}(t) \rangle \nonumber \\
&= \lim_{t\rightarrow \infty} \frac{1}{2}\int_0^\beta d\lambda \Big [\int_{-t+i\lambda}^{t+i\lambda} dz   \langle \hat{Q}(0) \hat{P}(z) \rangle \Big]. \label{simpf_Mqp1}
\end{align}
The last step requires time-translation by $t+i\lambda$ and changing variable to $z\rightarrow t+i\lambda$.
We can now do the integration over $z$ using contour integration. For this, we choose a contour of the rectangle in complex-plane joining the points $(-t,i\lambda)$, $(t,i\lambda)$, $(t,0)$, $(-t,0)$. Assuming no singularities in the upper half plane, we have
\begin{align}
\label{contour}
&\int_{-t+i\lambda}^{t+i\lambda} dz   \langle \hat{Q}(0) \hat{P}(z) \rangle = \int_{-t}^{t} dt^\prime   \langle \hat{Q}(0) \hat{P}(t^\prime) \rangle \nonumber \\
&+ i\int_0^{\lambda}dy \Big[ \langle \hat{Q}(0)\hat{P}(t+iy) \rangle - \langle \hat{Q}(0)\hat{P}(-t+iy) \rangle \Big].
\end{align}
By Eq.~\ref{necessary}, the term in square brackets in Eq.~\ref{contour} vanishes as $t\rightarrow \infty$.  Hence, substituting in Eq.~\ref{simpf_Mqp1}, we get
\begin{align}
\label{simpf_Mqp2}
&M(\hat{Q},\hat{P}) = \frac{1}{2} \int_{-\infty}^{\infty} dt   \langle \hat{Q}(0) \hat{P}(t) \rangle \nonumber \\
& = \frac{1}{2} \int_{-\infty}^{\infty} dt   \langle \hat{Q}(-t) \hat{P}(0) \rangle \hspace{5pt} \textrm{(time-translation by $-t$)}\nonumber \\
& = \frac{1}{2} \int_{-\infty}^{\infty} dt   \langle \hat{Q}(t) \hat{P}(0) \rangle \hspace{5pt} \textrm{(change variable $t\rightarrow -t$)}.
\end{align}
Thus we recover the expression in Eq.~\ref{notation2}.

%

\section{Derivation of $\rho_{NESS}^{\hat{\mathcal{H}}}$}\label{Dyson}
Starting with $\rho_{NIS}^{\hat{\mathcal{H}}}$ (Eq.~\ref{non-eq_intitial}), $\rho_{NESS}^{\hat{\mathcal{H}}}$ was obtained by observing $\rho_{NIS}^{{\hat{\mathcal{H}}}}=\rho_{EIS}^{{\hat{\mathcal{H}}}^\prime}$, with ${\hat{\mathcal{H}}}^\prime={\hat{\mathcal{H}}}+\epsilon {\hat{\mathcal{H}}}_{pert}$ (Eq.~\ref{H_perturb}), and $\frac{\partial\rho}{\partial t}= i [\rho,{\hat{\mathcal{H}}}] = i [\rho,{\hat{\mathcal{H}}}^{\prime}] - i\epsilon [\rho,{\hat{\mathcal{H}}}_{pert}]$. Dyson series of standard time-dependent perturbation theory was used to obtain $\rho_{NESS}^{\hat{\mathcal{H}}}$. Here we give the steps in detail.

First, we go to interaction picture with respect to ${\hat{\mathcal{H}}}^{\prime}$.
\begin{align}
&\rho^I(t) = e^{i{\hat{\mathcal{H}}}^{\prime} t}\rho(t)e^{-i{\hat{\mathcal{H}}}^{\prime} t}, \nonumber \\
&{\hat{\mathcal{H}}}_{pert}^I(t) = e^{i{\hat{\mathcal{H}}}^{\prime} t}{\hat{\mathcal{H}}}_{pert} e^{-i{\hat{\mathcal{H}}}^{\prime} t}.
\end{align}
Then, we have $\frac{\partial\rho^I}{\partial t}= - i\epsilon [\rho^I(t),{\hat{\mathcal{H}}}_{pert}^I(t)]$, which gives
\begin{align}
&\rho^I(t) = \rho^I(0) -i\epsilon \int_0^t dt^\prime [\rho^I(t^\prime),{\hat{\mathcal{H}}}_{pert}^I(t^\prime)]\simeq \rho^I(0) \nonumber \\
&-i\epsilon \int_0^t dt^\prime [\rho^I(0),{\hat{\mathcal{H}}}_{pert}^I(t^\prime)].
\end{align}
To obtain the second line we have used the first line recursively in the RHS and have kept only terms upto linear order in $\epsilon$. Going back to Schroedinger picture, and recalling $\rho^I(0)=\rho_{EIS}^{{\hat{\mathcal{H}}}^\prime}$, we get
\begin{align}
&\rho(t) \simeq e^{-i{\hat{\mathcal{H}}}^{\prime} t}\rho_{EIS}^{{\hat{\mathcal{H}}}^\prime}e^{i{\hat{\mathcal{H}}}^{\prime} t} \nonumber \\
&-i\epsilon \int_0^t dt^\prime [e^{-i{\hat{\mathcal{H}}}^{\prime} t}\rho_{EIS}^{{\hat{\mathcal{H}}}^\prime}e^{i{\hat{\mathcal{H}}}^{\prime} t},e^{-i{\hat{\mathcal{H}}}^{\prime} t^\prime}{\hat{\mathcal{H}}}_{pert}e^{i{\hat{\mathcal{H}}}^{\prime} t^\prime}].
\end{align}
Now, taking $t\rightarrow \infty$, and then taking ${\hat{\mathcal{H}}}^\prime \rightarrow {\hat{\mathcal{H}}}$ in the second term noting that corrections above this will be of higher order in $\epsilon$, we have our desired equation for $\rho_{NESS}^{\hat{\mathcal{H}}}$ (Eq.~\ref{rho_NESS}).

\section{Kubo trick}\label{Kubo_trick}

Here we give the steps for the simplification in Eq.~\ref{O_NESS}. This involves a standard technique used in deriving Kubo formula, which we call the Kubo trick.
We have $Tr(\hat{O}[{\hat{\mathcal{H}}}_{pert}(-t),\rho])=Tr(\hat{O}(t)[{\hat{\mathcal{H}}}_{pert},\rho])=\langle [\hat{O}(t), {\hat{\mathcal{H}}}_{pert}]\rangle$ by time-translation invariance.
Let $\hat{K}=({\hat{\mathcal{H}}}-\mu \hat{N})$. Then,
\begin{align}
\label{simp2}
&[{\hat{\mathcal{H}}}_{pert},\rho] = [{\hat{\mathcal{H}}}_{pert},\frac{e^{-\beta \hat{K}}}{Z}]=\rho \tilde{\Phi}(\beta) \nonumber \\
&\tilde{\Phi}(\lambda) = e^{\lambda \hat{K}} {\hat{\mathcal{H}}}_{pert} e^{-\lambda \hat{K}} - {\hat{\mathcal{H}}}_{pert}
\end{align}
Thus,
\begin{align}
& \frac{d\tilde{\Phi}(\lambda)}{d\lambda} = e^{\lambda \hat{K}} [\hat{K},{\hat{\mathcal{H}}}_{pert}] e^{-\lambda \hat{K}} = e^{\lambda {\hat{\mathcal{H}}}} [{\hat{\mathcal{H}}},{\hat{\mathcal{H}}}_{pert}] e^{-\lambda {\hat{\mathcal{H}}}} \nonumber \\
&=-i e^{\lambda {\hat{\mathcal{H}}}}\frac{d{\hat{\mathcal{H}}}_{pert}}{dt}e^{-\lambda {\hat{\mathcal{H}}}} = -i \dot{{\hat{\mathcal{H}}}}_{pert}(-i\lambda)
\end{align}
where we have used $[{\hat{\mathcal{H}}}_{pert},N]=0$ and $\dot{{\hat{\mathcal{H}}}}_{pert}\equiv \frac{d{\hat{\mathcal{H}}}_{pert}}{dt}=-i[{\hat{\mathcal{H}}}_{pert},{\hat{\mathcal{H}}}]$.  Then, we have,
\begin{align}
\label{simp3}
\tilde{\Phi}(\beta) = -i\int_0^\beta d\lambda \dot{{\hat{\mathcal{H}}}}_{pert}(-i\lambda) 
\end{align}
Using Eq.~\ref{simp2}, Eq.~\ref{simp3}, we have
\begin{align}
&\langle [\hat{O}(t), {\hat{\mathcal{H}}}_{pert}]\rangle = -i\int_0^\beta d\lambda Tr(\hat{O}(t) \rho \dot{{\hat{\mathcal{H}}}}_{pert}(-i\lambda)) \nonumber \\
&= -i\int_0^\beta d\lambda \langle \dot{{\hat{\mathcal{H}}}}_{pert}(-i\lambda) \hat{O}(t) \rangle
\end{align}  
Using above equation and Eqs.~\ref{curr_def}, \ref{H_perturb} and \ref{notation}, we get the second line in Eq.~\ref{O_NESS}.

\section{Current correlations to density correlations \label{Green-Kubo}}
Here we give the proof of Eq.~\ref{isolated_thermodynamic}. We want to look at
\begin{align}
\label{sigma_GK3}
&\lim_{L\rightarrow \infty}\mathcal{K}^S(L,t)= \lim_{L \rightarrow \infty} \frac{\beta}{2(L-1)}\int_{-t}^t dt^\prime \langle\langle \hat{I}_S(t^\prime) \hat{I}_S(0) \rangle\rangle_S  \nonumber \\
& = \lim_{L\rightarrow \infty} \frac{\beta}{L-1} \int_{0}^{t} dt^\prime \textrm{Re}\left(\langle \langle \hat{I}_S(t^\prime) \hat{I}_S(0) \rangle\rangle_S\right).
\end{align}
In going from the first line to the second line, we have used time translation by $t\rightarrow -t$, and the fact that $\hat{I}_S$ is Hermitian so, $\left(\hat{I}_S(t) \hat{I}_S(0) \right)^\dagger=\hat{I}_S(0) \hat{I}_S(t)$. Now we use the continuity equations for a `local' Hamiltonian,
\begin{align}
\hat{N}_S=\sum_{p=-\infty}^\infty \hat{n}_p,~~\frac{d \hat{n}_p}{dt} = \hat{I}_{p-1} - \hat{I}_{p},~~ 
\hat{I}_S=\sum_{p=-\infty}^{\infty} \hat{I}_p.
\end{align}
Here, we have already assumed thermodynamic limit and neglected the boundary terms. Then we observe that
\begin{align}
&\frac{d}{dt_1}\frac{d}{dt_2} \left[\sum_{p,q=-\infty}^\infty (p-q)^2\langle \langle \hat{n}_p(t_1) \hat{n}_q(t_2)\rangle\rangle_S \right] \nonumber \\
&= \sum_{p,q=-\infty}^\infty (p-q)^2 \langle\langle \left(\hat{I}_{p-1}(t_1)-\hat{I}_{p}(t_1)\right)\left(\hat{I}_{q-1}(t_2)-\hat{I}_{q}(t_2)\right)\rangle\rangle_S \nonumber \\
& = -2 \sum_{p,q=-\infty}^\infty \langle\langle \hat{I}_{p}(t_1)\hat{I}_{q}(t_2)\rangle\rangle_S. 
\end{align}
To arrive at the last line, we have shifted the dummy indicies $p$ and $q$ such that $\langle\langle \hat{I}_{p}(t_1)\hat{I}_{q}(t_2)\rangle\rangle_S$ can be factored out. We define $\tau=t_1-t_2$, then $\frac{d}{dt_1}=\frac{d}{d\tau}$, $\frac{d}{dt_2}=-\frac{d}{d\tau}$. Using time translation symmetry, this gives,
\begin{align}
\label{m2nnII}
&\frac{d^2}{d\tau^2}\left[\sum_{p,q=-\infty}^\infty (p-q)^2\langle \langle \hat{n}_p(\tau) \hat{n}_q(0)\rangle\rangle_S \right] \nonumber \\
&= 2 \sum_{p,q=-\infty}^\infty \langle\langle \hat{I}_{p}(\tau)\hat{I}_{q}(0)\rangle\rangle_S \nonumber \\
&\Rightarrow \frac{d}{d\tau}\left[\sum_{p,q=-\infty}^\infty (p-q)^2\langle\langle \hat{n}_p(\tau) \hat{n}_q(0)\rangle\rangle_S \right] \nonumber \\
&=2 \int_0^{\tau} dt \langle\langle\hat{I}_{S}(t)\hat{I}_{S}(0)\rangle\rangle_S.
\end{align}
Using Eq.~\ref{m2nnII} and ~\ref{sigma_GK3}, we find Eq.~\ref{isolated_thermodynamic}. Eq.~\ref{main_eqn} for the open system is obtained by following exactly the same steps, but without taking the thermodynamic limit first, and carefully keeping the boundary terms.

\section{Finding transmission function}\label{transmission}

\begin{figure}
\includegraphics[width=\columnwidth]{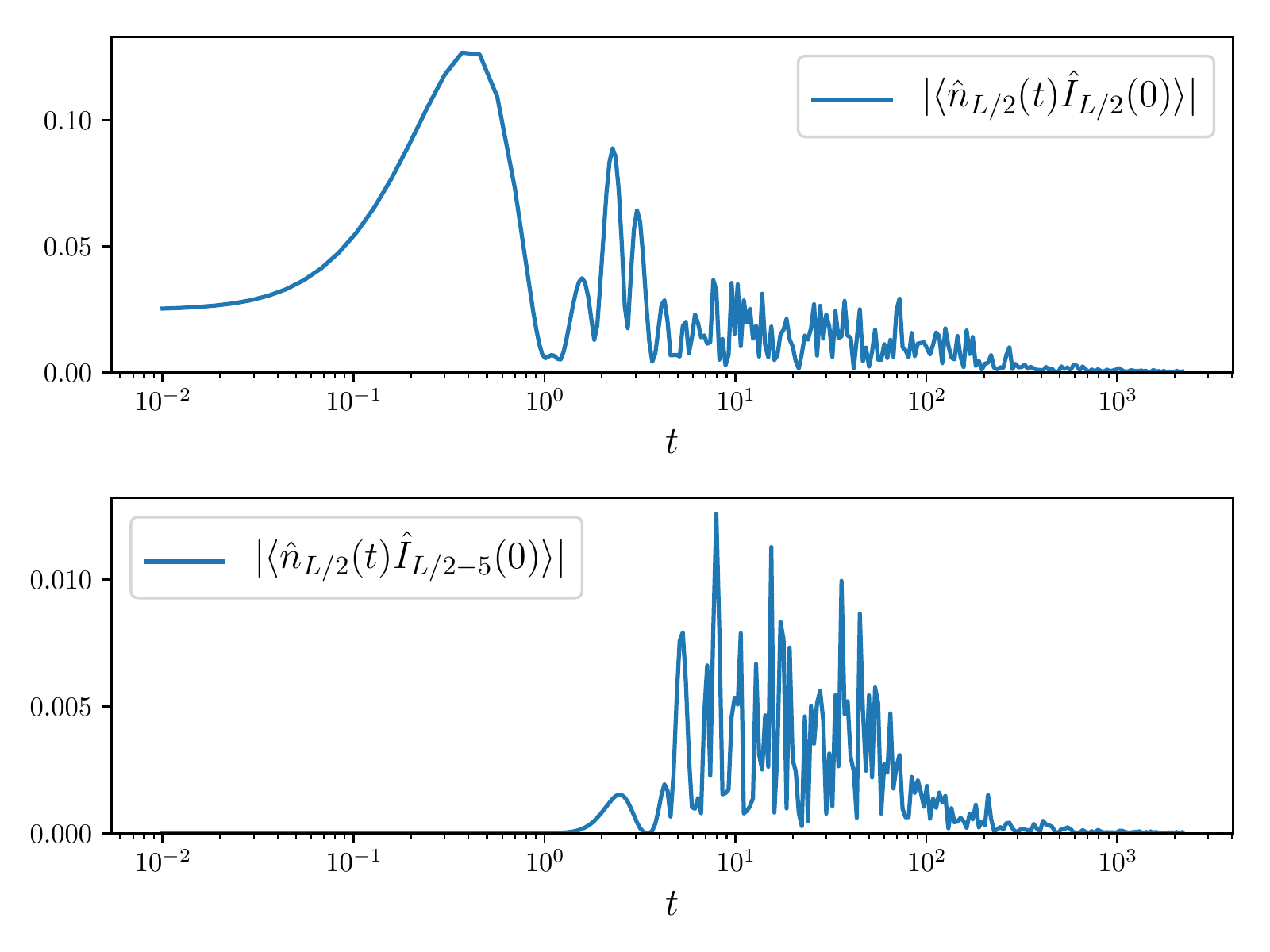}
\caption{(color online)  We explicitly check the validity of the mixing assumption for local particle currents and densities (see Eq.~\ref{mixing}) of the open critical AAH model. Both $\langle \hat{n}_{L/2}(t)\hat{I}_{L/2}(0) \rangle$ (top) and  $\langle \hat{n}_{L/2}(t)\hat{I}_{L/2-5}(0) \rangle$ (bottom) goes to zero with time, which is consistent with the assumption. Parameters: $L=21$, bath length $L_B=3307$, $\beta=0.1$, $\mu=1$, $\gamma=1.5$, $t_B=1.5$.}
\label{fig:AAH_check_mixing_open}
\end{figure}

\begin{figure}
\includegraphics[width=\columnwidth]{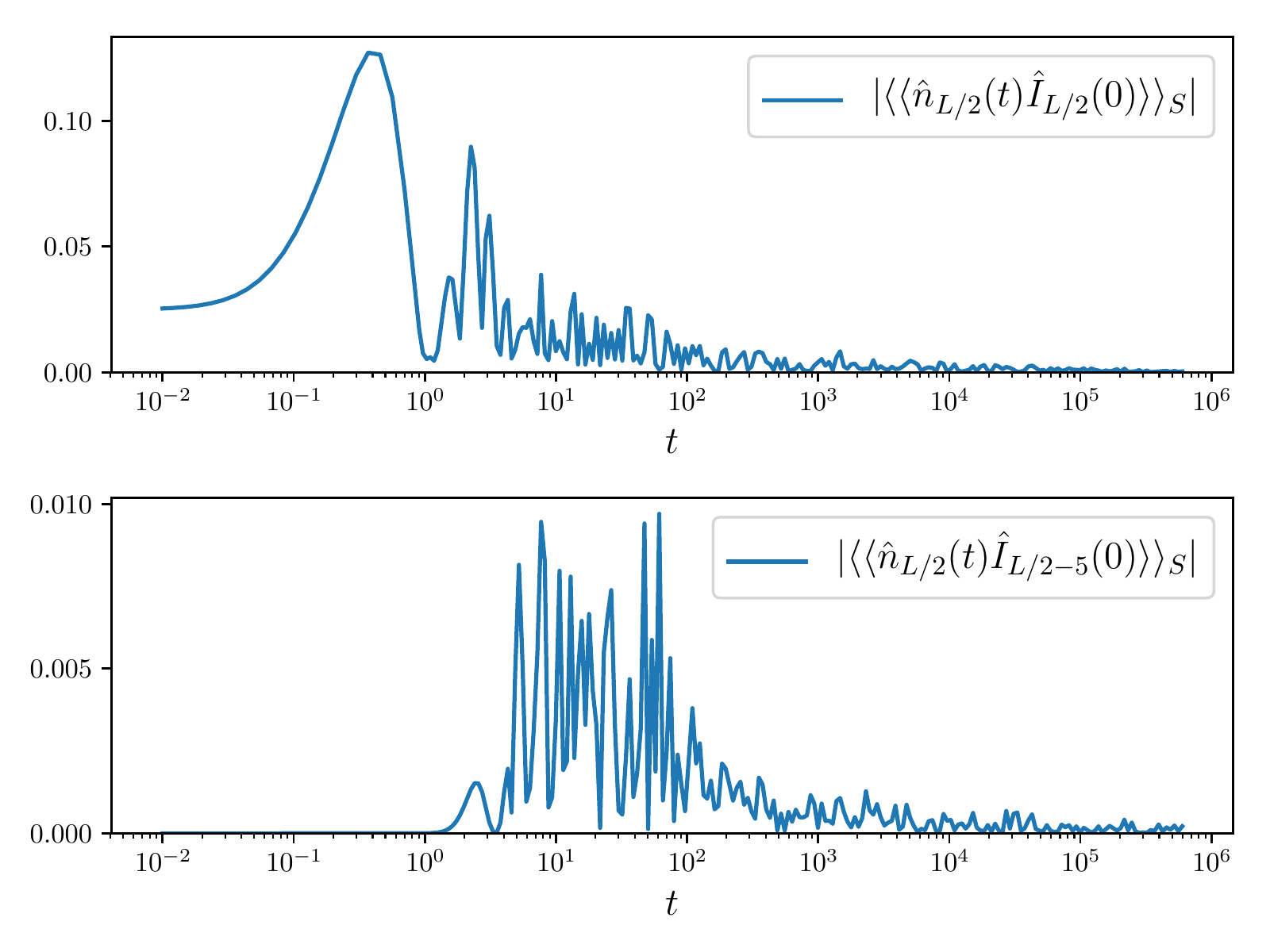}
\caption{(color online)  For comparison, we plot $\langle \langle \hat{n}_{L/2}(t)\hat{I}_{L/2}(0) \rangle \rangle_S$ (top) and  $\langle\langle \hat{n}_{L/2}(t)\hat{I}_{L/2-5}(0) \rangle\rangle_S$ (bottom), which are the isolated critical AAH model quantities corresponding the ones plotted in Fig.~\ref{fig:AAH_check_mixing_open} for the open system.  It is apparent that equivalents of Eq.~\ref{mixing} is also valid for the isolated critical AAH model in thermodynamic limit. Parameters: $L=10946$, $\beta=0.1$, $\mu=1$.}
\label{fig:AAH_check_mixing_isolated}
\end{figure}

We can write any non-interacting (quadratic) system Hamiltonain in the form $\hat{\mathcal{H}}_S=\sum_{i,j=1}^L \hat{c}_i^\dagger [\mathbf{H}_S]_{ij} \hat{c}_j$.
Let $\mathbf{G}(\omega) = \mathbf{M}^{-1}(\omega)$ be the non-equilibrium Green's function (NEGF) of the set-up. $\mathbf{M}(\omega)$ is given by the $L\times L$ matrix $\mathbf{M}(\omega)~=~\left[ \omega\mathbf{I}- \mathbf{H}_S - \mathbf{\Sigma}^{(1)} (\omega)-\mathbf{\Sigma}^{(L)} (\omega)\right]$, where $\mathbf{\Sigma}^{(1)} (\omega)$, $\mathbf{\Sigma}^{(N)} (\omega)$ are   bath self energy matrices with the only non-zero elements given by  
\begin{align}
\mathbf{\Sigma}^{(p)}_{pp}(\omega) = -\mathcal{P}\int_{-2t_B}^{2t_B} \frac{d\omega^\prime \mathfrak{J}(\omega^\prime)}{2\pi(\omega^\prime-\omega)}-\frac{i}{2}\mathfrak{J}(\omega), \hspace{5pt} p=1,L
\end{align}
where $\mathcal{P}$ denotes principal value. $\mathfrak{J}(\omega)$ is the bath spectral function. For our model of baths in Eq.~\ref{baths}, the bath spectral function is given by
\begin{align}
\mathfrak{J}(\omega)=\frac{2\gamma^2}{t_B}\sqrt{1-(\frac{\omega}{2t_B})^2}.
\end{align}
The transmission function is given by 
\begin{align}
T(\omega) = \mathfrak{J}^2(\omega) \mid \mathbf{G}_{1N}(\omega)\mid^2= \frac{\mathfrak{J}^2(\omega)}{\mid\textrm{det}\left[\mathbf{M}\right]\mid^2}.
\end{align}

\section{Checking mixing assumption for open critical AAH model}\label{check_mixing}

We have shown in Fig.~\ref{fig:AAH_all}(d) that $\mathcal{K}^O(L,t)$ indeed reaches a steady state value given by $(L-1)G$. Our derivation shows that for this to be valid the mixing assumption for local currents and densities (Eq.~\ref{mixing}) has to be valid. In Fig.~\ref{fig:AAH_check_mixing_open}, we explicitly check this for our set-up defined in Eqs.~\ref{H_S_AAH}, \ref{baths} (see Fig.~\ref{fig:AAH_set_up}). Indeed, as expected from Eq.~\ref{mixing}, $\langle \hat{n}_{m}(t)\hat{I}_{\ell}(0) \rangle$ goes to zero with increase in time. We show this by explicitly plotting $|\langle \hat{n}_{L/2}(t)\hat{I}_{L/2}(0) \rangle|$ and $|\langle \hat{n}_{L/2}(t)\hat{I}_{L/2-5}(0) \rangle|$. 

Though not directly required for our theory, just for comparison, we also plot $|\langle\langle \hat{n}_{L/2}(t)\hat{I}_{L/2}(0) \rangle\rangle_S|$ and $|\langle\langle \hat{n}_{L/2}(t)\hat{I}_{L/2-5}(0) \rangle\rangle_S|$ in Fig.~\ref{fig:AAH_check_mixing_isolated}. As defined in Eq.~\ref{K_m_def}, $\langle\langle ... \rangle \rangle_S$ denotes that the average is taken over the system thermal state $\rho_S = e^{-\beta({\hat{\mathcal{H}}}_S-\mu \hat{N}_S)}/Tr(e^{-\beta({\hat{\mathcal{H}}}_S-\mu \hat{N}_S)})$ and the time translation operator  involves only ${\hat{\mathcal{H}}}_S$. Here, ${\hat{\mathcal{H}}}_S$  is the critical AAH model Hamiltonian defined in Eq.~\ref{H_S_AAH}. Although the decay takes much longer time, it is apparent that for isolated critical AAH model in thermodynamic limit, the equivalent of Eq.~\ref{mixing} also holds.

\section{Transport in critical AAH model}\label{AAH_transport}

Transport in isolated critical AAH model is not actually strictly diffusive. As shown in Ref.~\cite{oc_mismatch3}, the higher moments show a crossover from diffusive to super-diffusive scaling at long time. But, the time scale required to see this crossover in $m_2^{S}(t)$ is so large that it cannot be seen in within our current computational abilities. Nevertheless, the sub-diffusive behavior seen in the open system is not expected to show up. Moreover, the point here is, within time scales and the system sizes possible to explore with current computational abilities, $m_2^{S}(t)$ shows almost perfect diffusive scaling while $G$ shows almost perfect sub-diffusive scaling. 

\bibliography{ref_synopsis1}
\end{document}